\documentclass[sigconf]{acmart}

\usepackage{booktabs} 
\usepackage[ruled,algo2e]{algorithm2e} 
\usepackage{algorithm}
\usepackage{algpseudocode} 
\usepackage{graphicx}
\usepackage[labelfont=bf,textfont=small]{subcaption}
\usepackage{tabularx}
\usepackage{epsfig}   
\usepackage{xcolor}
\usepackage{amsthm}
\usepackage{multirow}
\usepackage{appendix}
\usepackage{xcolor}
\usepackage{pdfpages}
\usepackage{url}
\usepackage{soul}
\newcommand{\colortext}[1]{{\color{magenta}{#1}}}

\newcommand{\eat}[1]{}

\theoremstyle{definition}

\SetAlFnt{\small}
\SetAlCapFnt{\small}
\SetAlCapNameFnt{\small}
\SetAlCapHSkip{0pt}
\IncMargin{-\parindent}
\SetKwInput{kwInit}{Initialize}

\newenvironment{narrowFigure}[1][]{\begin{figure}[#1]}{\vspace{-0.4cm}\end{figure}}
\newenvironment{narrowTable}[1][]{\begin{table}[#1]}{\vspace{-0.6cm}\end{table}}
\usepackage{xcolor,etoolbox}
\usepackage[version=4]{mhchem}

\copyrightyear{2020}
\acmYear{2020}
\acmConference[WWW '20]{Proceedings of The Web Conference 2020}{April 20--24, 2020}{Taipei, Taiwan}
\acmBooktitle{Proceedings of The Web Conference 2020 (WWW '20), April 20--24, 2020, Taipei, Taiwan}
\acmPrice{}
\acmDOI{10.1145/3366423.3380108}
\acmISBN{978-1-4503-7023-3/20/04}

\makeatletter
\newcommand*{\inlineequation}[2][]{%
  \begingroup
    \refstepcounter{equation}%
    \ifx\\#1\\%
    \else
      \label{#1}%
    \fi
	\relpenalty=0
	\binoppenalty=0
    \ensuremath{%
      #2%
    }%
    ~\@eqnnum
  \endgroup
}
\makeatother
\makeatletter
\definecolor{SAEblue}{rgb}{0, .62, .91}
\renewcommand{\eqref}[1]{\textup{\eqreftagform@{\ref{#1}}}}
\let\eqreftagform@\tagform@
\def\tagform@#1{%
  \maketag@@@{\color{SAEblue}(\ignorespaces#1\unskip\@@italiccorr)}%
}
\makeatother

\makeatletter
\let\@msm@th@eqref\eqref
\renewcommand{\eqref}[1]{%
  \begingroup
  \leavevmode
  \color{SAEblue}%
  \hypersetup{linkbordercolor=[named]{SAEblue}}%
  \@msm@th@eqref{#1}%
  \endgroup
}
\makeatother

\begin{document}

\title[Cross-modal Identity Leakage between Biometrics and Devices]{Nowhere to Hide: \\Cross-modal Identity Leakage between Biometrics and Devices}

\author{Chris Xiaoxuan Lu}
\authornote{Also affiliated with the University of Oxford.}
\affiliation{%
  \institution{University of Liverpool}
}
\email{luxiaoxuan555@hotmail.com}

\author{Yang Li}
\affiliation{%
  \institution{New York University}
}
\email{zjzsliyang@nyu.edu}

\author{Yuanbo Xiangli}
\authornote{Corresponding author.}
\affiliation{%
  \institution{The Chinese University of Hong Kong}
}
\email{xy019@ie.cuhk.edu.hk}

\author{Zhengxiong Li}
\affiliation{%
  \institution{University at Buffalo, SUNY}
}
\email{zhengxio@buffalo.edu}

\begin{abstract}

Along with the benefits of Internet of Things (IoT) come potential privacy risks, since billions of the connected devices are granted permission to track information about their users and communicate it to other parties over the Internet. Of particular interest to the adversary is the user identity which constantly plays an important role in launching attacks.
While the exposure of a certain type of physical biometrics or device identity is extensively studied, the compound effect of leakage from both sides remains unknown in multi-modal sensing environments.
In this work, we explore the feasibility of the compound identity leakage across cyber-physical spaces and unveil that co-located smart device IDs (e.g., smartphone MAC addresses) and physical biometrics (e.g., facial/vocal samples) are side channels to each other. 
It is demonstrated that our method is robust to various observation noise in the wild and an attacker can comprehensively profile victims in multi-dimension with nearly zero analysis effort.
Two real-world experiments on different biometrics and device IDs show that the presented approach can compromise more than $70\%$ of device IDs and harvests multiple biometric clusters with $\sim94\%$ purity at the same time. 


\end{abstract}



\thanks{}

\maketitle


\section{Introduction} 
\label{sec:introduction}

Internet of Things (IoT) devices are gaining popularity: it is expected that $20$ billion IoT devices will be globally deployed by $2020$ \cite{meulen2017gartner}. Along with the benefits of them come potential privacy risks. By connecting rich sensors (e.g., cameras, microphones, geophones) to the Internet, many IoT devices can potentially \emph{suggest or expose} extensive information about their users. Moreover, due to the mature manufacturing technologies, IoT devices can be made very small and easily hidden in commonplace \cite{hidden_cam_legal}. According to the Identity Theft Research Center, $\sim 4.5$ million sensitive records containing personally identifiable information were exposed in 2017 alone, representing a $126\%$ increase from the previous year \cite{identity_theft}. Users who have their identity information compromised may suffer from threats ranging from persecution by governments \cite{marthews2017government}, to targeted frauds/advertising \cite{christin2010dissecting}, to spoofing attacks \cite{hadid2015biometrics}. 

Given the severe consequences of identity theft, prior efforts have been made to investigate leak channels and effective countermeasures. These efforts can be broadly categorized into two types, where one focuses on revealing the identity of user via biometrics (e.g., face, voice, gait), while the other type concerns the link between user identity and devices identities (IDs), such as (MAC addresses of smartphones, web/browser cookies of laptops etc.). For biometrics, Ilia et al. proposed to rethink the access control of face photos to prevent users being recognized by unwanted individuals on social media \cite{ilia2015face}. Meanwhile, privacy breach via device IDs also draws substantial attention. For example, Yen et al. showed how several digital identities, such as IP, cookies and usernames, can be combined to track users' devices reliably on the web \cite{yen2012host}. Cunche et al. demonstrated that the MAC addresses of Wi-Fi enabled devices can be abused to track device owners' locations \cite{cunche2014know}.

The above works unveil possible compromisations brought by either biometrics or device IDs. However, few studies discuss the risk of information leakage from a compound channel. Intuitively, given a hidden camera contributing user's facial biometrics, the device IDs (e.g., phone MAC addresses \cite{cunche2014know}) captured by a \emph{co-located} WiFi sniffer can be utilized as \emph{side information} to complete its eavesdropping view. Similarly, a co-located hidden camera is also a side channel to the WiFi sniffer and can be maliciously used to augment its knowledge base. 
Though the tactic to combine side channels is straightforward, associating one's biometrics and devices is non-trivial under real-world constraints. The first challenge is the \textbf{spatial-temporal mismatch} between biometric and device ID observations.
While both of the observations conceptually describe the `identity' of a user, they are not necessarily sensed at the same time. For instance, sniffing the MAC address of a victim's smartphone does not imply that the victim is speaking at the exact same instant and vice versa. In addition, device IDs and biometrics are usually sensed by different sensors with different spatial coverage. This difference can further exacerbate the mismatch issue. For example, as radio wave can penetrate obstacles, a WiFi sniffer is able to sniff devices behind a shelf which occludes the cameras for face capture. Such spatial mismatch is inevitable when the attacker seeks to eavesdrop, where the field covered by \emph{hidden/spy} cameras or microphones becomes more constrained \cite{wifi_spy_mic}. To further complicate the problem, the spatial-temporal mismatch cannot be easily addressed by naive elimination, because eavesdropped data often contains \textbf{unexpected disturbances from the crowd} which cause a substantial amount of observations that are unimportant to the attacker. Concretely, in our experiment, $1,147$ facial images from $35$ people outside the target (victim) group were observed (see Sec.~\ref{sub:data_collection_methodology}). Many of these non-targets were temporal visitors to the eavesdropping space. Additionally, we also found $109,768$ \emph{distinct} MAC addresses in our $238,833,555$ sniffed packets, with only $27$ MAC addresses belonged to targets' device IDs. As a result, a naive elimination method is insufficient to handle these unexpected disturbances, while a more sophisticated method is required to associate biometrics and device IDs. 


In this paper, we first validate the feasibility of identity breach across biometrics and device IDs. We then investigate a cross-modal association method that addresses the aformentioned challenges an attacker might face. Our method draws the intuition that: 
\textbf{even if sensed at different instants, long-term eavesdropped biometrics and device IDs still form a shared context across modalities}. 
Given the shared context, an adversary is able to assign a set of MAC addresses to biometric clusters based on their attendance consistency throughout eavesdropped sessions. To deal with real-world complexity, we present a method that allows an attacker to robustly associate cross-modal identities under observation mismatches and substantial noises. Our contributions are both conceptual and empirical:
\begin{itemize}
	\item We observe a new privacy issue of cross-modal identity leakage and formalize an unprecedented threat in multi-modal sensing environments. Our work unveils a compound identity leak from the combined side channels between human biometrics and device identities. 

	\item We present a novel approach that robustly associates physical biometrics with device IDs under substantial sensing noises and observation disturbances. If maliciously used, an attacker can automatically profile the victims' identities in multiple dimensions at the same time.


	\item We conduct extensive evaluation on two real-world data collection of different biometrics, showing that our presented attack is a real threat. In certain cases, an attacker is able to correctly de-anonymize over $70\%$ device IDs and harvest multiple biometric clusters of $\sim94\%$ purity.

\end{itemize}

Our prototype and code are available at \colortext{\url{https://github.com/zjzsliyang/CrossLeak}}.

\section{Background} 
\label{sec:background}

\subsection{WiFi MAC Address and Sniffing} 
\label{sub:wifi_mac_address_sniffing}

\noindent \textbf{WiFi MAC Addresses}
WiFi communication is on the $802.11$ radio between end devices and access points. Similar to the IP address in network, every $802.11$ radio has a $48$-bit MAC address in link layer which is globally unique as the identifier of that device. The uniqueness of MAC addresses across devices is guaranteed by the Institute of Electrical and Electronics Engineers (IEEE) and the organizations such as smartphone manufacturer and retailer. 

\noindent \textbf{WiFi Sniffing.}
The information exchange on a network following IEEE $802$ standards is through packets. A packet is a chunk of data consists of `header', `body' and `trailer'. Particularly, `header' is where the sender address and destination address are stored, which provides the most crucial information for attackers who are interested in identification. The communication between two devices on WiFi is by convention via packets but not necessarily in an one-on-one form. Instead, the sender broadcasts the packet on the airwaves where every device within range can `listen' to it. The packet is then checked by any receiving node, and compared with their own MAC addresses. As a result, besides the desired device whose MAC address is specified in the header, `header' information is also exposed to the surrounding ones which causes passive sniffing.



\begin{narrowFigure}[t]\centering
\includegraphics[width=\columnwidth]{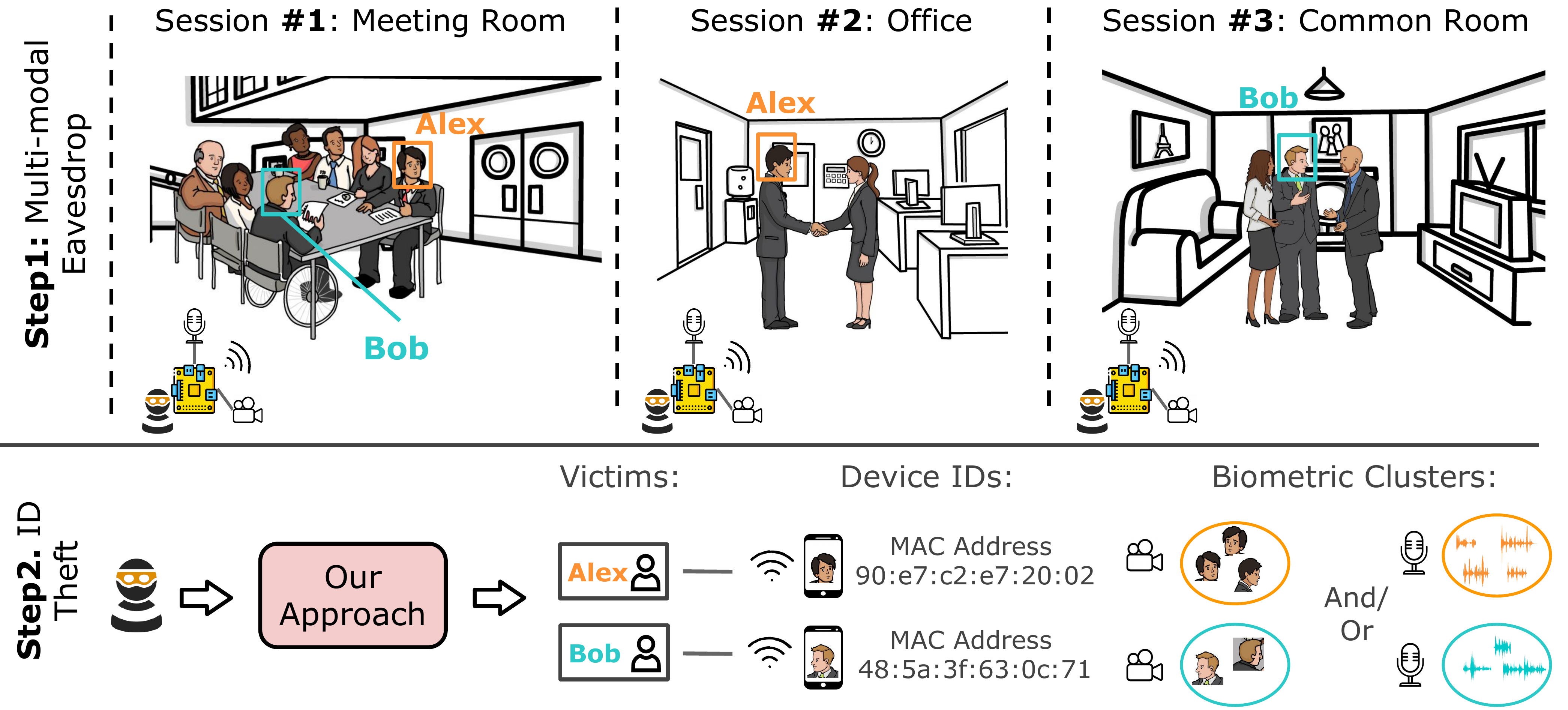}  
\caption{Attack Scenario. Through multi-modal eavesdropping (e.g., via a hidden camera and/or microphone integrated with a WiFi sniffer), an attacker stealthily collects biometric and device data in multiple sessions. The attacker can then leverage our presented approach to automatically \emph{identify} targets' (e.g., Alex and Bob in the figure) device IDs and \emph{harvest} their biometric clusters. Subjects other than Alex and Bob are \emph{out-of-set} subjects (i.e. non-targets).}
\label{fig:attack_scenario}
\end{narrowFigure}

\subsection{Threat Model} 
\label{sub:threat_model}

The adversary model considered in this work is an attacker somehow eavesdropped the device IDs and biometrics of victims for a period of time. The eavesdropping tool can be a combination of a WiFi/Bluetooth/ZigBee sniffer and hidden microphone/camera, or an integrated solution customized from existing devices (e.g., WiFi spy microphone \cite{wifi_spy_mic}). These tools are cheap and commercially available, allowing an easy setup for attackers. 
Deploying such an eavesdropping tool is feasible, with threats coming from both insiders and outsiders.

\noindent \textbf{Insider Threat.}
A threat from insiders has the potential to compromise a group of other co-workers' identities (see Fig.~\ref{fig:attack_scenario}). Owning to the privileged access granted to an insider, deploying the eavesdropping tool is easy to disguise. The attacker can place it as an office decoration without arousing suspicion. Moreover, since such a setup does not generate packets, it will not be detected even if the wireless traffics in the target workplace is regularly inspected by a security guard.
In fact, recent studies \cite{hidden_cam_legal} found that disguised eavesdropping on a particular type of biometrics or device IDs already exists in domestic and commercial environments. Under this assumption, the adversary can stealthily gather device identities (e.g., MAC addresses) and physical biometrics (e.g., faces/voices). When sufficient information is collected, the attacker can use our presented approach for bilateral ID association. As visually or acoustically identifying a co-worker is easy for the adversary, such association builds a bridge for the attacker to figure out who owns which device and enables further attacks (e.g., online or physical tracking \cite{korolova2018cross,castelluccia2012behavioural,cunche2014know}).

\noindent \textbf{Outsider Threat.}
Even if long-term sensor deployment is inaccessible to outsiders, an adversary can still eavesdrop enough information by carrying disguised \emph{mobile devices} to public spaces.  
Here we provide a concrete example with a WiFi sniffer and camera. As WiFi sniffing is available on many laptops, we imagine that an attacker uses his laptop (probably equipped with a back camera) as the eavesdropping system. The attacker then brings his laptop and sits in the corner of a public place (e.g., a cafe or vehicles with WiFi services). Using this setup, the adversary is able to continuously and covertly capture high-profile or normal customers' facial images and sniff WiFi packets of surrounding user smartphones for days. Eventually, the adversary can associate individuals' facial images to personal device IDs with our presented approach. The multi-dimensional IDs of the victims can make a profit by selling them to third-party stakeholders, or can be used for launching personalized advertising/fraud \cite{kshetri2019online} and biometric spoofing \cite{rodrigues2010evaluation} subsequently. 

\section{Dataset and Feasibility} 
\label{sec:dataset_and_feasibility}

To understand how one can associate device IDs to biometrics, we designed an eavesdropping prototype to collect multi-modal data in indoor environments. We then study the feasibility of associating these two types of identities on the collected datasets that provides insights for the presented approach.

\begin{narrowFigure}[!t]
\centering
  \begin{subfigure}{0.48\linewidth}
      \centering
      \includegraphics[width=\linewidth]{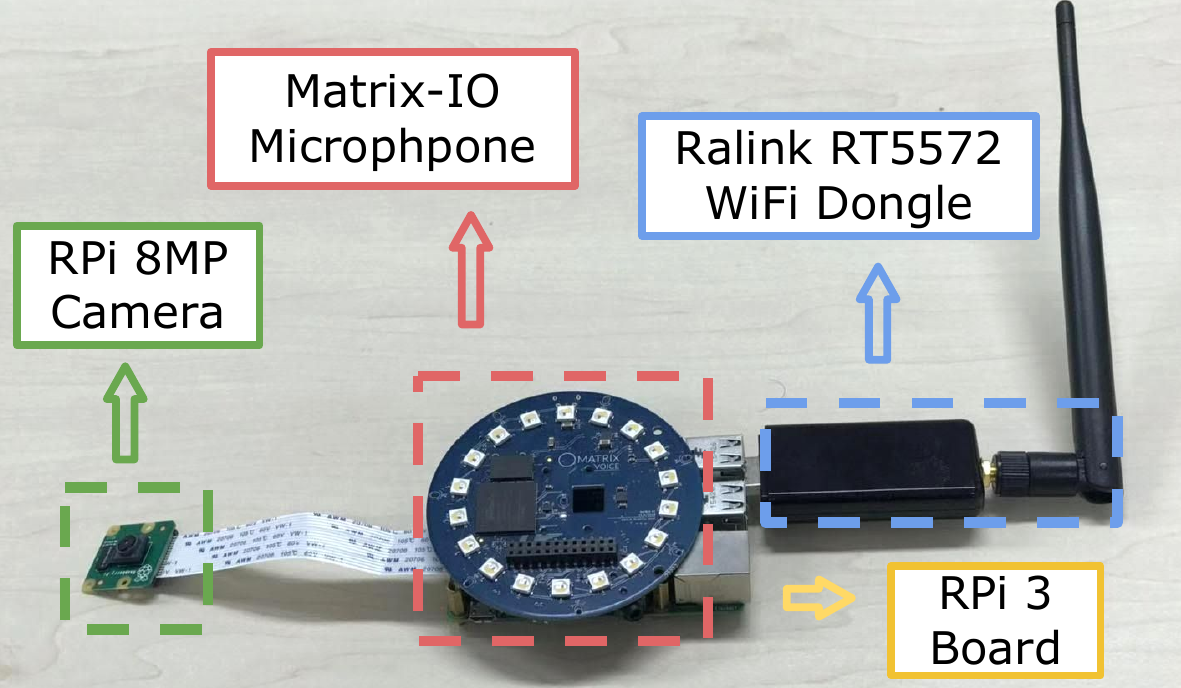}
      \caption{Prototype}
      \label{fig:prototype}
    \end{subfigure}%
 	\hfill
    \begin{subfigure}{0.5\linewidth}
      \centering
      \centering
      \includegraphics[width=\linewidth]{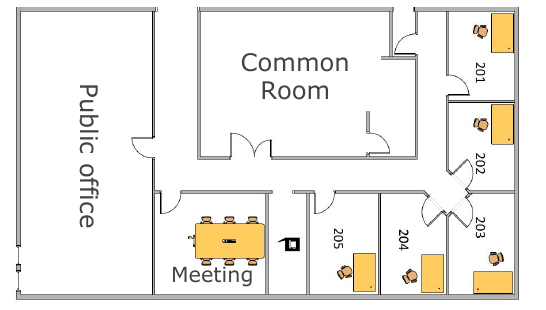}
      \caption{Testbed}
      \label{fig:testbed}
    \end{subfigure}%
    \caption{Data collection. (\ref{fig:prototype}): assembled eavesdropping prototype; (\ref{fig:testbed}): floor plan of the eavesdropping testbed; data were collected in its three shared spaces - public office, common and meeting rooms.}
    \label{fig:eaves_setup}
\end{narrowFigure}

\subsection{Data Collection Methodology} 
\label{sub:data_collection_methodology}
 


\noindent \textbf{Eavesdropping Prototype.} 
As shown in Fig.~\ref{fig:prototype}, the assembled eavesdropping prototype is built on a Raspberry Pi (RPi) 3 \cite{rpi3}. To collect facial/vocal samples and WiFi MAC addresses, the sensing modules consist of (1) a Matrix-IO board \cite{matrix_io} to record audios in the far field, (2) an RPi 8MP Camera Board \cite{rpi_camera} to capture videos and (3) an Ralink RT5572 WiFi Dongle \cite{ralink_5572} to sniff MAC addresses. 
To reduce storage overheads, the system records video/audio input to a circular buffer, while simultaneously analyze the input to detect motion or voice activity on the RPi. When motion or voice activity is detected, the buffer is saved to a file. 
On the side of sniffing, RPi runs TShark in the background and continuously captures packets transmitted on the airwaves. The RPi then extracts the source MAC address from each packet and saves it along with the time of it being captured and its corresponding received signal strength in a file.
In order to eavesdrop as many MAC addresses as possible, channel hopping is launched along with sniffing such that live packets on all channels of different WiFi bands can be captured. It is worth mentioning that as an alternative for integrated eavesdropping, the information can also be collected from separate devices as long as they are co-located. 

\noindent \textbf{Testbed and Ethics.} As shown in Fig.~\ref{fig:testbed}, our testbed is a certain floor in a commercial building that has $32$ regular residents. A subset of those residents volunteered to participate as \emph{victims} in our experiment.
We deployed three eavesdropping tools in a \emph{public office}, a \emph{common room} and a \emph{meeting room}, with areas of $60m^2$, $46m^2$ and $25m^2$ respectively. Our system unobtrusively collected \emph{victims}' data from their daily activities, which is referred as \emph{sessions} (e.g., meetings, chats etc.) in our context\footnote{The study has received ethical approval \emph{R50950}.}. Sessions being accidentally recorded without full consents were deleted permanently.

\noindent \textbf{Face-domain Dataset.} 
This dataset consists of $123$ sessions and there are $22$ victims and $35$ non-victims (e.g., temporary visitors) captured by the camera. After face detection and image pre-procssing, the video clips give $27,482$ facial samples. Over $2 \times 10^8$ packets are sniffed during eavesdropping, where approximately $110,000$ of them are distinct. Please refer to Tab.\ref{tab:dataset_description} for more detail.

\noindent \textbf{Voice-domain Dataset.} 
This dataset contains $49$ sessions contributed by $21$ victims and $9$ non-victims through different conversation scenarios (e.g., meetings). The collected audio recordings are segmented into $3,555$ utterances. In total, there are around $2 \times 10^6$ packets sniffed during data collection coming from about $3,500$ distinct MAC addresses. Please refer to Tab.\ref{tab:dataset_description} for more detail.

\begin{narrowTable}[!t]
\centering
\small
  \begin{tabular}{|c|c|c|}
  \hline
   & \textbf{Face Dataset} & \textbf{Voice Dataset} \\ \hline
  Sessions & 123 & 49 \\ \hline
  Victims (Volunteers) & 22 & 21 \\ \hline
  Non-Victims & 35 & 9 \\ \hline
  Biometric Samples & 27,482 & 3,555 \\ \hline
  Sniffed Packets & 238,833,555 & 1,985,862 \\ \hline
  Distinct Device IDs & 109,755 & 3,478 \\ \hline
  \begin{tabular}[c]{@{}c@{}}Average Victims / Session\end{tabular} & 5.90 & 2.14 \\ \hline
  \begin{tabular}[c]{@{}c@{}}Average Duration / Session\end{tabular} & 2 hr & 15 min \\ \hline
  \end{tabular}
  \caption{Key Features of Our Dataset.}
  \label{tab:dataset_description}
\end{narrowTable}

\subsection{Feasibility Exploration} 
\label{sub:feasibility}

We are now in a position to explore the feasibility by inspecting the eavesdropped data. For the ease of illustration, we use the term \emph{session} hereafter to broadly refer to settings in which victims interact with entities in an environment for a given time interval, e.g., a physical visit or a meeting in the meeting room from 10am to 11am. 
Therefore, a dataset comprises a set of sessions, each of which is uniquely determined by the time-slot, the location and the involved subjects.

\begin{narrowFigure}[!t]
    \centering
    \begin{subfigure}{\linewidth}
        \centering
        \includegraphics[width=.493\linewidth]{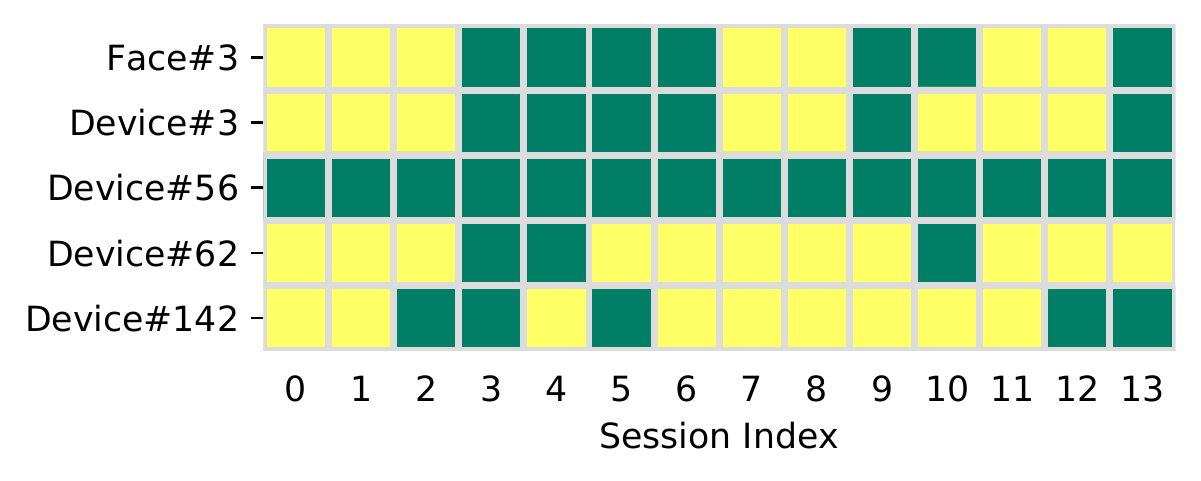}
        \hfill
        \includegraphics[width=.493\linewidth]{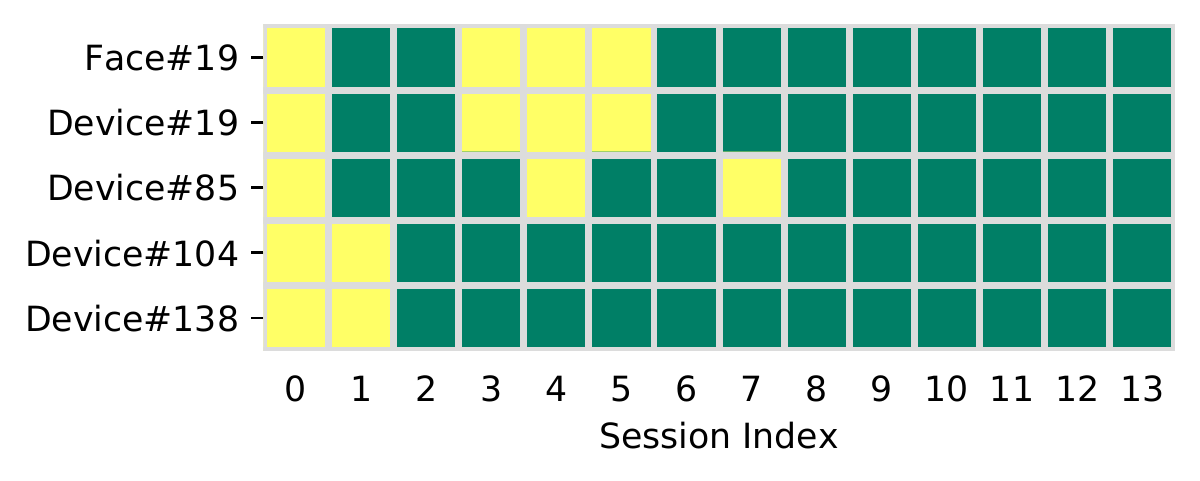}
        \caption{Face Cluster $\rightarrow$ Device ID: Left - User \#3; Right - User\#19}
        \label{fig:face2device}
    \end{subfigure}
    \vskip\baselineskip
    \begin{subfigure}{\linewidth}
        \centering
        \includegraphics[width=.493\linewidth]{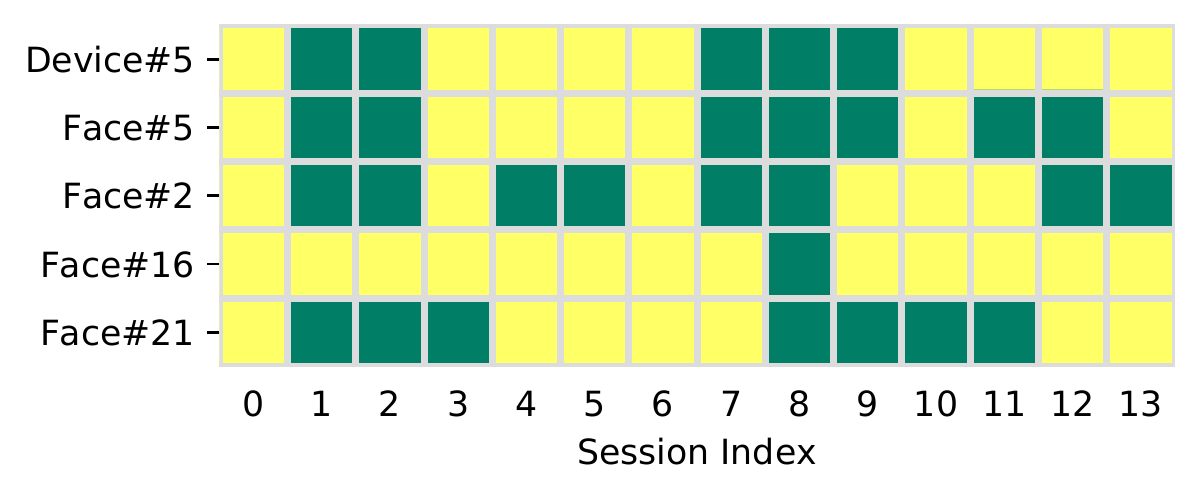}
        \hfill
        \includegraphics[width=.493\linewidth]{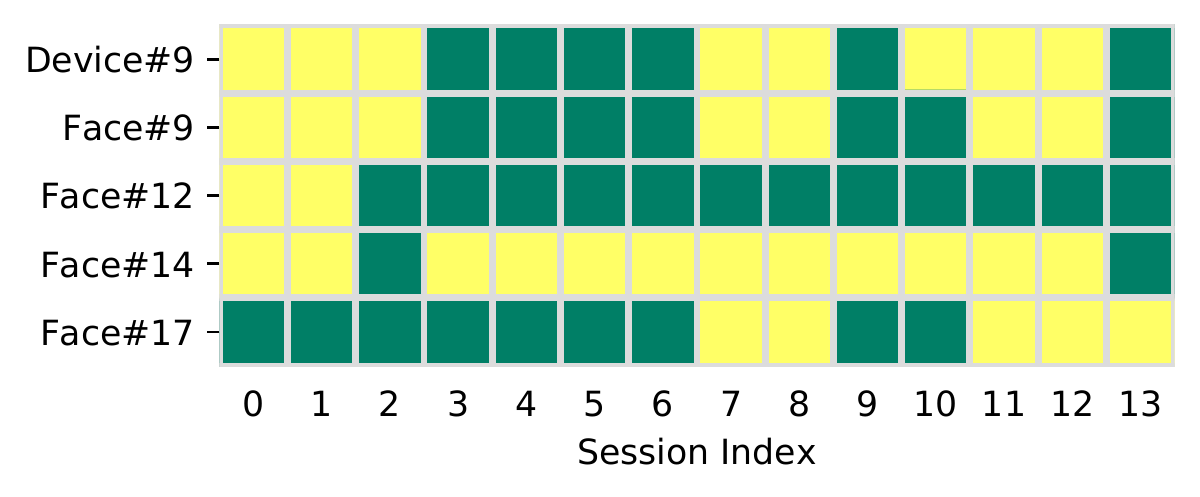}
        \caption{Device ID $\rightarrow$ Face Cluster: Left - User \#5; Right - User\#9}
        \label{fig:device_id2face}
    \end{subfigure}
    \caption{Attendance map of $14$ sessions: dark squares indicate absence and bright squares indicate presence. From the compatibility of attendance in (\ref{fig:face2device}), one can identify which device belongs to a target face cluster and vice versa in (\ref{fig:device_id2face}).}
    \label{fig:attendance_map}
\end{narrowFigure}

\subsubsection{How Biometrics Aid the De-Anonymization of MAC addresses?}
\label{ssub:bio-deanonymize}

Take facial biometric as an example, we firstly perform clustering on the eavesdropped facial images according to the similarity between their deep features, similar to \cite{lu2019autonomous,xiangli2019iscan}. Note that because of the intrinsic uncertainty, the derived clusters are impure, mixed with samples of other victims or non-targets (to be further discussed in Sec.~\ref{sub:a_naive_method_and_its_limitation_}). 
Fig.~\ref{fig:face2device} shows the attendance of $2$ randomly selected victims derived from their facial biometric clusters observed in $14$ random sessions. Below the map of face clusters, we show the top-$4$ device IDs with the most similar attendance patterns. One can notice that, the session attendance map of the `true' smartphone has a very similar pattern with the victim's face cluster. As a result, if the face cluster is visually identifiable by the attacker (e.g., an insider threat), such similarity can be maliciously used to deanonymize the MAC address of a target device. 

\subsubsection{How MAC Addresses Aid the Harvest of Biometrics?} 

Equally, an attacker can use the eavesdropped MAC addresses to improve the efficacy of biometrics harvesting. It is inevitable that eavesdropped biometrics in real-world include erroneous samples coming from non-victims or non-targets (e.g., short term visitors to a compromised office as demonstrated in Fig.~\ref{fig:attack_scenario}). These samples can not only contaminate the victims' biometric clusters, but result in irrelevant clusters that mislead the attacker. Even worse, if the eavesdropping is conducted in a public space (e.g., cafe in a commercial building) by an outsider threat, the amount of irrelevant clusters can be non-negligible. For example, in our face-domain dataset, there are $1,147$ facial images belongs to $35$ non-victims.
Despite these factors, as shown in Fig.~\ref{fig:attendance_map}, the face clusters' attendance maps of non-targets are much different from the target devices (i.e., cluster $\#5$ and $\#9$ in Fig.~\ref{fig:device_id2face}). The attacker can therefore focus on the successfully associated cluster-to-MAC address pairs and ignore the unassociated ones of non-targets.

\subsubsection{Uniqueness Analysis} 
\label{ssub:uniqueness_analysis}

Knowing that there exists session attendance similarity between one's physical biometrics and his/her personal device, a natural question to ask is, does every user have a \emph{discriminative attendance pattern} such that the target can be isolated? To answer this, we investigate the possibility of distinguishing a group of victims by selecting only a small subset of sessions in their attendance history. Two different sampling strategies are used to obtain the subset: (1) randomly selecting $G$ attended sessions of the victim (Rand-G) and (2) randomly selecting $G$ \emph{continuous} sessions attended by the victim (Cont-G). Under these strategies, we calculate the percentage of users that can be distinguished from others. Fig.~\ref{fig:quan_fea} shows that on the face-domain dataset, $18$ out of the $22$ victims can be distinguished from the randomly selected $25$ sessions (Rand-$25$ strategy); when the number of sessions reaches $50$, all victims can be perfectly distinguished. Similar behaviour can be observed on face-domain dataset when adopting Cont-G strategy. As for the voice-domain dataset which has $49$ sessions, $19$ out of the $21$ victims can be identified via Rand-$20$ and Cont-$20$ strategies; when all sessions are used, all victims are perfectly separated. \\

\noindent \emph{\textbf{Takeaway:}}
The above results indicate that the attendance pattern of a victim can be significantly different from others. Such uniqueness probably is due to the distinct living/working habits of individuals. 
Based on the above analysis, we conclude that cross-modal identity breach is feasible and is a real threat.

\begin{narrowFigure}[!t]
\centering
  \begin{subfigure}{0.49\linewidth}
      \centering
      \includegraphics[width=\linewidth]{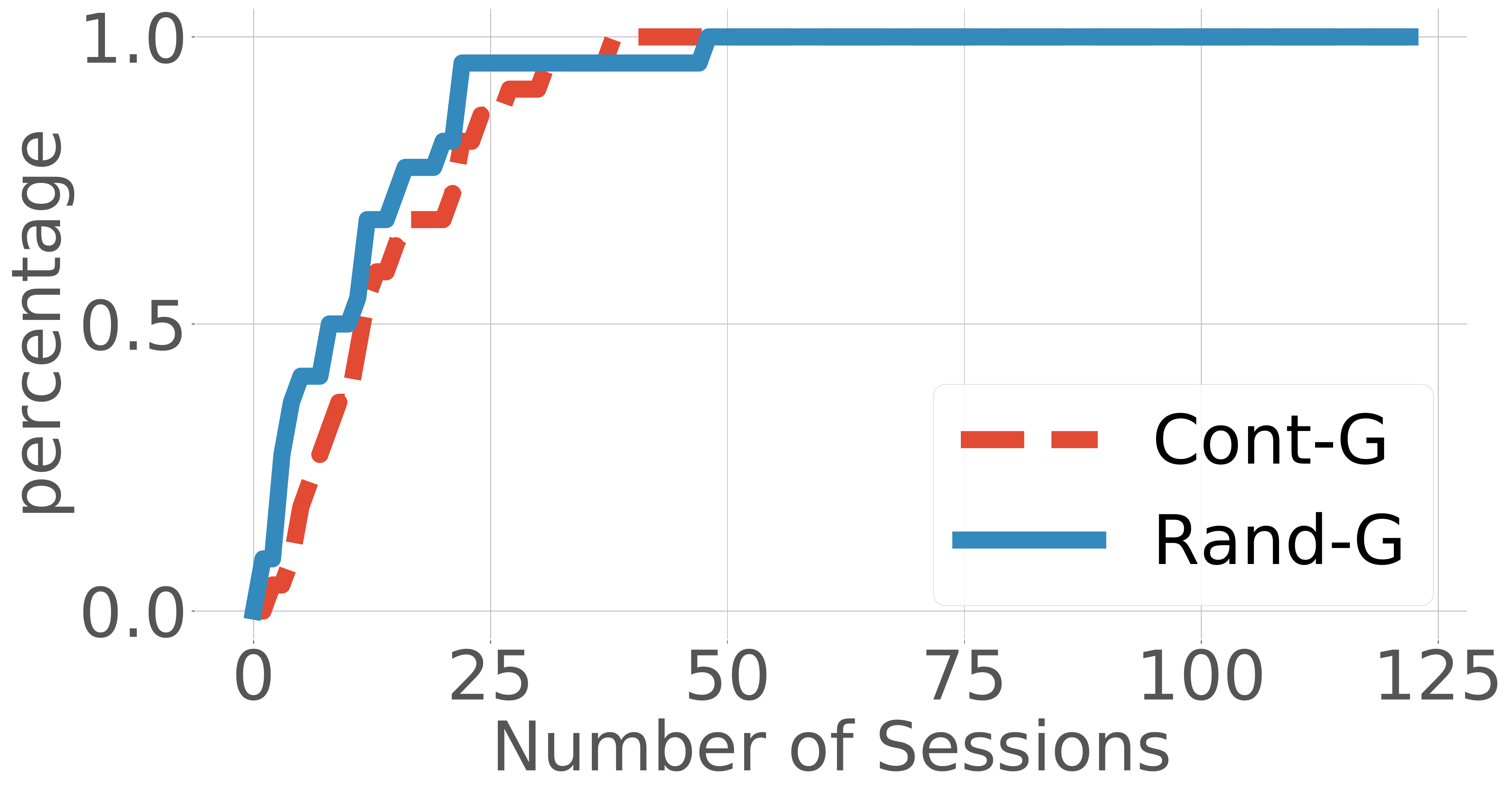}
      \caption{Face-domain Dataset}
    \end{subfigure}%
    \hfill
    \begin{subfigure}{0.49\linewidth}
      \centering
      \centering
      \includegraphics[width=\linewidth]{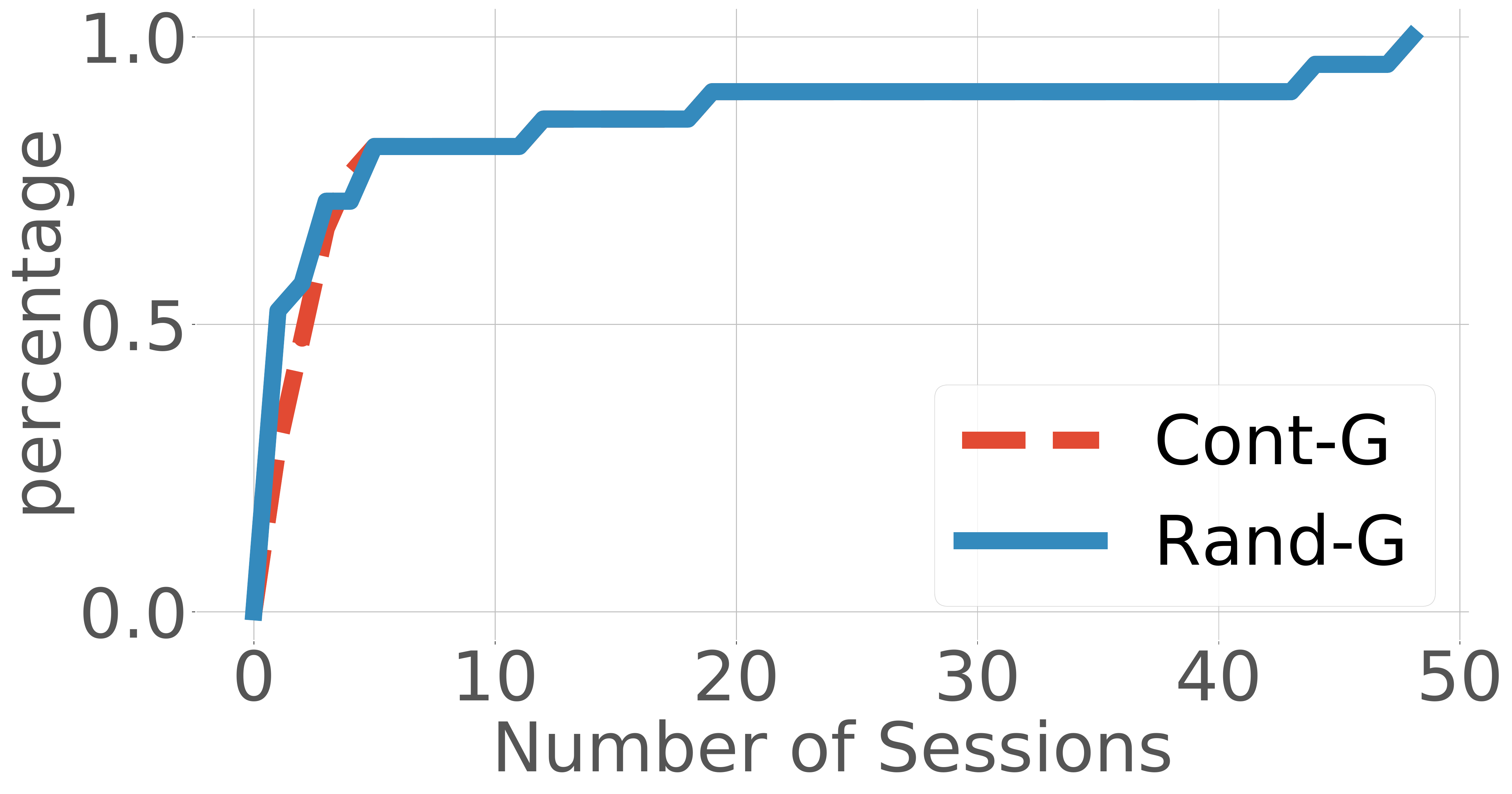}
      \caption{Voice-domain Dataset}
    \end{subfigure}%
    \caption{Quantified Feasibility: percentage of distinguishable users based on different session sampling strategies.}
    \label{fig:quan_fea}
\end{narrowFigure}

\section{Overview and Formulation} 
\label{sec:attack_overview}

\subsection{Attack Overview} 
\label{sub:overview}

We now present an attack vector that leverages the above intuition to associate victims' device IDs and physical biometrics. Our approach features its robustness to eavesdropping disturbances, requiring minimal prior knowledge from an adversary to launch the attack. As shown in Fig.~\ref{fig:attack_overview}, our framework consists of three modules that operates in a pipelined manner: (1) Multi-modal eavesdropping (2) Device filtering and (3) Cross-modal ID association. 

The \emph{multi-modal eavesdropping} module described in Sec.~\ref{sub:data_collection_methodology} is used to gather data. An attacker then feeds the sniffed device IDs to the \emph{device filtering} module so that a majority of devices unimportant to the attacker can be removed. Given the filtered device IDs and biometric collection, the attacker can associate them by tasking the \emph{cross-modal association} module. In what follows, we will first formulate the generic association problem after eavesdropping and then introduce the filtering and association modules in Sec.~\ref{sec:device_filtering} and Sec.~\ref{sec:cross_modal_id_association} respectively.


\subsection{Problem Formulation} 
\label{sub:problem_formulation}

\noindent \textbf{Definition.}
Without loss of generality, suppose the adversary aims to compromise the identities of $P$ victims in a target environment. The adversary can either be an insider that attacks co-workers or an outsider who collects data of people in public spaces. 
We assume the scenario where the attacker managed to eavesdrop identities in $G$ sessions, $\mathcal{S} = \{s_j|j=1,2, \ldots, G\}$. The term \emph{session} is introduced in Sec.~\ref{sub:feasibility} that broadly refers to settings in which victims interact with entities in an environment for a certain amount of time.
The set of eavesdropped device IDs are denoted as $\mathcal{Y} = \{y_j|j=1,2, \ldots, H\}$. A set of biometric samples are also captured, denoted as $\mathcal{X} = \{x_j|j=1,2, \ldots, W\}$. 

\noindent \textbf{Modeling Real-world Disturbances.}
In real-world scenarios, it is possible that eavesdropping will experience unexpected disturbance from the out-of-set (OOS) observations, including device IDs and biometrics outside the target group. For instance, a subject visiting the vulnerable workspace in a short term without connecting to WiFi will only have their facial images captured. Similarly, since WiFi signals are able to penetrate walls, the sniffed MAC addresses are likely to include devices from a long distance, while the owner has never appeared in the camera. Without imposing strong assumptions, we consider the case in the wild where OOS observations occur in both eavesdropping $\mathcal{X}$ and $\mathcal{Y}$. Aside from the OOS disturbances, a large portion of the sniffed MAC addresses are dummies or noises, owning to the MAC address randomization mechanism \cite{martin2017study} and nuisance packets generated by other WiFi devices, e.g., access points, workstations etc. Our problem setup considers that these observation noises are included in $\mathcal{Y}$.

\noindent \textbf{Attack Goal.} \emph{Given the very noisy eavesdropped data of $\mathcal{X}$ and $\mathcal{Y}$, discover the device IDs and biometric feature clusters belonging to $P$ target victims, and then correctly associate these cross-modal identities}. For readability, we describe our presented approach in the context using MAC addresses as the device IDs with facial/vocal samples as biometrics. 

\begin{narrowFigure}[t]\centering
\includegraphics[width=\columnwidth]{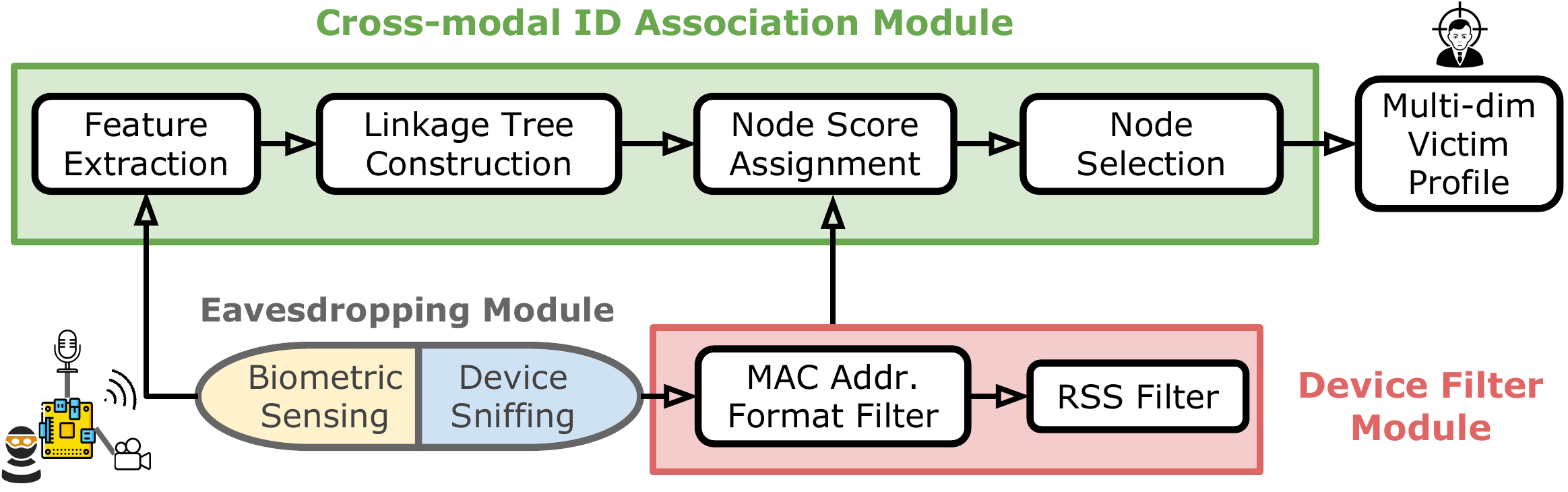}  
\caption{Attack Overview. Our attack framework consists of three modules that operate in a pipelined manner: (1) Multi-modal eavesdropping (2) Device filtering and (3) Cross-modal ID association.}\label{fig:attack_overview}
\end{narrowFigure}

\section{Device Filtering} 
\label{sec:device_filtering}

As the sniffed MAC addresses (i.e., device IDs) contain instances outside the target environment, one needs to filter the eavesdropped device IDs in $\mathcal{Y}$ and derive a smaller and cleaner set of MAC addresses $\mathcal{L} = \{l_j|j=1,2, \ldots, M\}$. Our proposed filter draws characteristics from (1) MAC address format and (2) wireless signal strength. It should be noted that while we describe the following filtering approach with the case of WiFi MAC addresses, the similar concept can generalize to device IDs attained by other means, e.g., Bluetooth sniffing \cite{albazrqaoe2016practical,bluetooth_sniffing}.

\subsection{MAC Address Format Filter}
\label{sub:filter_based_on_mac_address_format}

The goal of this sub-module is to remove device IDs that do not comply with end devices' (e.g., smartphones) formats by inspecting the information hidden in MAC addresses. 
Particularly, we focus on two dominant types of disturbance: (1) Randomized MAC addresses and (2) MAC addresses of WiFi access points. 

\subsubsection{MAC Randomization} 
\label{ssub:subsubsection_name}

To prevent WiFi tracking, both Android and Apple iOS operating systems allow devices in a \emph{disassociated state} to use random, locally assigned MAC addresses in active WiFi scanning. These addresses are dummy bytes pointing to no real devices.
The MAC addresses remain random until the device is associated with an access point (AP). However, abiding by the 802.11 protocol of IEEE, one very specific bit of the MAC address is the seventh bit of its first byte: the Locally Administered (LA) bit. If a LA bit set to $1$, the MAC address is implied as randomized/changed by the Administrator of the device \cite{WifiNigel2018}. Therefore, the first step of our filter utilizes this LA bit and inspects each sniffed packet. Only those MAC addresses identified as non-randomization are passed to next-step analysis. As we can see in Tab.~\ref{tab:mac_filter}, randomized MAC addresses account for about $65\%$ and $20\%$ in the distinct sniffed device IDs on our face and voice data collection.

\subsubsection{Vendor Information} 
\label{ssub:vendor_information}

Aside from the information about randomization, one can also find the vendor information from a MAC address. To guarantee the uniqueness of MAC addresses across devices, the IEEE assigns blocks of addresses to organizations in exchange for a fee. An Organizationally Unique Identifier (OUI), also the first three bytes of a non-randomized MAC address, is such an information snippet that may be purchased and registered with the IEEE \cite{OUI}. By looking up the public database of OUI, we first find the manufacturers of all MAC addresses. As the attacker targets the mobile devices of victims, the filter removes those addresses whose manufacturers are corresponding to WiFi access points and switches (e.g., TP-Link, Cisco, 3Com, Juniper, LinkSys, D-Link, NetGear). Tab.~\ref{tab:mac_filter} lists that $91$ and $96$ MAC addresses in our face and voice data collection are such kind of MAC addresses. 

\begin{narrowTable}[!t]
\small
\centering
	\begin{tabular}{|c|c|c|c|c|}
	\hline
	\multirow{2}{*}{\textbf{\begin{tabular}[c]{@{}c@{}}Filter\\ Module\end{tabular}}} & \multicolumn{2}{c|}{\textbf{Face Dataset}} & \multicolumn{2}{c|}{\textbf{Voice Dataset}} \\ \cline{2-5} 
	 & Number & Percentage & Number & Percentage \\ \hline
	Random & 71,720 & 65\% & 693 & 20\% \\ \hline
	Vendor & 91 & 0.08\% & 96 & 3\% \\ \hline
	RSS & 108,639 & 99\% & 3,378 & 97\% \\ \hline
	\end{tabular}
	\caption{Statistics of MAC addresses removed by different filter modules. NOTE: one nuisance MAC address is possibly detected by multiple modules, leading to over $100\%$ total percentage in the table.}
\label{tab:mac_filter}
\end{narrowTable}

\subsection{Received Signal Strength (RSS) Filter} 
\label{sub:received_signal_strength_filter}

The second sub-module of our filter aims to filter out distant devices outside the effective eavesdropping range. Here, we turn to the WiFi Received Signal Strength (RSS) information in sniffed packets. Following the signal propagation laws of wireless radios, RSS reflects \emph{how far} the sniffed devices are from the WiFi sniffer. This spatial information is pivotal to determining the collocation with biometric samples. Although WiFi sniffing can capture packets in distance ($\sim40$m), biometric sensing often works only in a short range (e.g., $7$m for microphones). In our context, MAC addresses of interest to attackers are those can be associated with co-located biometric observations; our filter thus discards those MAC addresses with low RSS values, e.g., less than $-45$ dBm on voice-domain dataset and less than $-55$ on face-domain dataset. This helps filter out $108,639$ and $3,378$ unimportant MAC addresses in our two datasets (see Tab.~\ref{tab:mac_filter}). The threshold is derived empirically by following the geo-fence concept introduced in \cite{lu2016robust}. In Sec.~\ref{subsub:sensitivity_rss}, we will discuss the impacts of different RSS thresholds on the overall performance. \\

\noindent \emph{\textbf{Summary:}}
Applying all above filtering sub-modules substantially narrows down the association range of device IDs. In our case, we found that the MAC addresses of interest (i.e., $\mathcal{L}$) is reduced from $109,768$ to $174$ on face datasets and from $3,478$ to $54$ on the voice dataset. This reduction makes the follow-up cross-modal association tractable and efficient.

\section{Cross-modal ID Association} 
\label{sec:cross_modal_id_association}

Given the filtered MAC addresses $\mathcal{L}$ and biometric samples $\mathcal{X}$, the next step is to discover the MAC addresses of victims and associate them across two identity spaces. Our association approach consists of three sequential processes: (1) biometric feature extraction (2) linkage tree construction and (3) node selection. At a high level, the algorithms works by first extracting deep features from biometric samples $\mathcal{X}$ and constructing a linkage tree based on their feature similarity. Each node in the tree is a biometric cluster. Then by leveraging the advocated concept of session attendance pattern (see Sec.~\ref{sub:feasibility}), victims' MAC addresses can be discovered in $\mathcal{L}$ and assigned to the corresponding biometric clusters. Algo.~\ref{algo:association} illustrates the association steps.

\begin{algorithm}
	\caption{\texttt{Cross-modality association.}}
	\label{algo:association}
	\textbf{Input: } pre-trained feature model $f_\theta$, biometric samples $\mathcal{X}$, device IDs $\mathcal{Y}$and estimated number of victims $K$.\\
    \textbf{Output: } decision variable $A = (a_{i,j})_{N*M}$ that victims' device ID $j$ with corresponding biometric cluster node $i$ are selected.
	\begin{algorithmic}[1]
	    \State $\mathcal{L} \gets \texttt{filtering\_device}(\mathcal{Y})$ \Comment{Sec.~\ref{sec:device_filtering}}
	    \State $\mathbf{Z} \gets f_\theta (\mathcal{X})$ \Comment{Feature Extraction in Sec.~\ref{sub:biometric_feature_extraction}}
	    \State $T \gets \texttt{linkage\_tree\_construction}(\mathbf{Z})$ \Comment{Sec.~\ref{sub:linkage_tree_construction}}
	    \State $r_l \gets \texttt{device\_context\_vector}(\mathcal{L})$
	    \State $r_t \gets \texttt{treenode\_context\_vector}(T)$          
	    \State $A \gets \texttt{node\_selection}(T, r_t, r_l, K)$ \Comment{Sec.~\ref{sub:node_selection}}
	\end{algorithmic}
\end{algorithm}

\subsection{Biometric Feature Extraction} 
\label{sub:biometric_feature_extraction}

The key to merging biometric samples lies in how to measure their similarity. For biometric samples such as facial images or vocal segments, accurately quantifying their similarity is challenging owning to multiple factors. For instance, face images can differ a lot due to lighting conditions \cite{li2007illumination}; voice quality may vary across HiFi microphones to low-cost microphones. Thanks to the recent advances in DNNs, it is proven that they are able to handle variable observation conditions through feature learning from massive amounts of training samples.  To utilize this advantage, biometric samples are first encoded into a feature embedding $\mathbf{Z}$ via a deep neural network $f_\theta$ designed for biometric recognition (e.g., facenet \cite{schroff2015facenet} for face recognition and x-vector \cite{snyder2018x} for speaker recognition). Such $f_\theta$ can be pre-trained on public datasets and learns an effective feature representation. 
The similarity between biometric samples are based on the extracted features with the pre-trained model.

\subsection{A Naive Method and Its Limitation.} 
\label{sub:a_naive_method_and_its_limitation_}

Given the similarity between biometric samples and filtered device IDs, a naive approach originated from the feasibility analysis (see Sec.~\ref{sub:feasibility}) is to leverage the diverse participatory information in multiple sessions and adopt a two-step association procedure: a) in the Clustering Step, biometric observations $\mathcal{X}$ are firstly grouped into clusters across all sessions, each of which potentially contains the biometric samples of a single victim; and then b) in the Data Association Step, the clusters are assigned with device IDs based on the similarity between their attendance patterns across eavesdropped sessions. Although this approach is simple and easy to implement, it is not robust against OOS observations. For example, a session may contain faces or voices of temporal visitors while their device ID is not captured. Due to the disturbance incurred by OSS subjects, the number of clusters is intractable as a priori for the attacker. A misleading clustering result, however, could further degrade the quality of data association. 

\subsection{Linkage Tree Construction} 
\label{sub:linkage_tree_construction}

To address the limitation of the naive method, we use a linkage tree to merge similar biometric samples in a hierarchy manner. Each node in the tree is a biometric cluster candidate, potentially belonging to a single victim. 


\noindent \textbf{Tree Structure.}
Based on the extracted features $\mathbf{Z}$, our presented algorithm compiles biometric samples into a linkage tree $T$. The leaf nodes $T_{leaf}$ are samples, while a branch node represents the cluster of all its descendant leaf nodes. Essentially $T$ represents the hierarchical clustering of all biometric observations in different sessions. Selecting a combination of nodes from the tree gives a specific clustering plan. Each node $t_i$ in $T$ is associated with a \emph{linkage score} $q_{link_i}$, describing the average feature similarity or compatibility between the data within the cluster it represents.

\noindent \textbf{Context Vector.}
A basic linkage tree can be developed as the above. However, due to the presence of domain differences, the features learned via a pre-trained DNN can deviate and mislead the node merging process.
For example, the learned feature representations on public datasets of front faces are not effective for the side profile of a face. When ineffective representation features are used, two face images of different subjects can be falsely merged into the same node which may cause unrecoverable knock-on effects on the ensuing association step.
In order to combat this problem, we augment the linkage tree by introducing context vectors to augment the knowledge of each node. A context vector of the MAC address is essentially a binary vector describing the session attendance. The vector's length is equal to the total number of eavesdropped sessions.
Concretely, let $\mathbf{r}_{t_i} = (r_{t_i}^1, r_{t_i}^2, \ldots, r_{t_i}^{G})$ be the context vector of a tree node $t_i$, where $G$ is the total number of the eavesdropped sessions.
$r_{t_i}^j$ is set to $1$ if node $t_i$ contains biometric samples eavesdropped from session $s_j$. 
Similarly, a MAC address $l_k$ is also linked with a context vector $\mathbf{r}_{l_j}$, and $\mathbf{r}_{l_j}^j$ is set to 1 only if $l_j$ is detected in session $s_j$. Intuitively, for a node $t_i$ and a MAC address $l_j$, if $\mathbf{r}_{t_i}$ and $\mathbf{r}_{l_j}$ are similar enough, it is very likely that the grouped samples under node $t_i$ are actually the biometric data of the subject who owns that MAC address $l_j$, as their presence patterns in sessions agree with each other the most.

\noindent \textbf{Node Scores and Dice Coefficient.}
Given the context vectors, we can additionally introduce a data association score to each of its node $t_i$, which represents the fitness of assigning an identity label to $t_i$ given the filtered MAC addresses $\mathcal{L}$. 
For a biometric node $t_i$, we define its data association scores with respect to $M$ filtered MAC addresses as a vector $\mathbf{q}_{assoc_i} = (q_{assoc_i}^1, q_{assoc_i}^2, \ldots, q_{assoc_i}^M)$, where the $j$-th score $q_{assoc_i}^j$ is the Dice coefficient \cite{duarte1999comparison} between the node context vector $\mathbf{r}_{t_i}$ and the device context vector $\mathbf{r}_{l_j}$. Intuitively, when matching with device IDs, the absence information is not as informative as the presence information, especially for \emph{users present rarely}. Dice coefficient is known to be suitable in handling such cases which favors the presence information (i.e., `1') when comparing two binary vectors. We will further explain this in Sec.~\ref{subsub:sensitivity_omega}.
Together with the linkage score, the final score assigned to the cluster node $t_i$ to a MAC address $l_j$ is a composite score function:
\begin{equation}\label{eq:score}
\small
  	q_i^j = (1-\omega) * q_{link_i} + \omega * q_{assoc_i}^j
 \end{equation}
where the parameter $\omega$ governs how much the adversary trusts the learned biometric features and to what extent the adversary wants them to impact the result of cross-modal association.

\subsection{Node Selection} 
\label{sub:node_selection}
\noindent \textbf{Optimization Program.}
Once node scores are all assigned, the attack of cross-modal ID association is equivalent to selecting the top $K$ nodes from the tree with a maximum sum. With the previously introduced terms and notations, we formulate this node selection problem as follows:


\begin{equation}
\small
	\max_{\mathbf{A}} \quad \sum^{N}\limits_{i=1} \sum^{M}\limits_{j=1} q_i^j * a_{i,j} \label{eq:obj} \\
\end{equation}
subject to \inlineequation[eq:ct_1]{\sum^{M}\limits_{j=1} a_{i,j} \leq 1, \forall i \in {1, \ldots, N}}, \inlineequation[eq:ct_2]{\sum^{N}\limits_{i=1} a_{i,j} \leq 1, \forall j \in {1, \ldots, M}}, \inlineequation[eq:ct_3]{\sum^{N}\limits_{i=1} \sum^{M}\limits_{j=1} a_{i, j} = K}, \inlineequation[eq:ct_4]{\sum\limits_{i \in \Pi_{k}} \sum^{M}\limits_{j=1} a_{i,j} \leq 1, \forall k \in T_{leaf}}, \inlineequation[eq:ct_5]{a_{i,j} \in \{0, 1\}, \forall i \in \{1, \ldots, N\}, \forall j \in \{1, \ldots,M\}}.

\noindent where $\mathbf{A} = (a_{i,j})_{N \times M}$ is the decision variable and $q_i^j$ is the composite score determined by Eq.~\eqref{eq:score}. $T_{leaf}$ represents the set of all leaf nodes in the linkage tree. The objective function aims to maximize the total scores when selecting $K$ nodes in the linkage tree $T$ with size of $N$. Intuitively, the selected $K$ nodes are the optimal clusters out of these $N$ biometric observations. The inequalities in Eq.~\eqref{eq:ct_1} simply mean that a node can be assigned to at most one device ID. Similarly, the constraints in Eq.~\eqref{eq:ct_2} are used to ensure each device ID is associated with a single node. Eq.~\eqref{eq:ct_3} decides how many nodes the program should select, i.e., the number of victims. 
As a clustering tree, a node cannot be selected with its ancestors or descendants at the same time since they contain duplicate data. In order to compile this tree structure in optimization, the constraint Eq.~\eqref{eq:ct_4} is enforced to guarantee that on any path leading to a leaf node, at most one node is assigned to a device ID. Finally, the constraint Eq.~\eqref{eq:ct_5} is there to make sure that decision variable $a_{i,j}$ can take on the integer value $0$ and $1$ only. 
The above optimization formulation is essentially an integer linear programming (ILP) problem and can be solved with off-the-shelf tools \cite{karmarkar1984new}.

\noindent \textbf{Choice of K.} 
To run the above optimization program, a key step is to estimate a rough number of subjects $K$ in the target environment. Such estimation is made based on the adversary's observation as either an insider or an outsider. However, due to the disturbance from the unknown number of OOS subjects, $K$ is often inaccurate  in practice but treated as an approximation of the true number of victims. Despite this, because the above association approach adopts a node selection strategy, it is able to \emph{pick out} and associate as many correct nodes as possible with either under- or over-estimated $K$, as illustrated in Fig.~\ref{fig:cross_assoc}.

\begin{narrowFigure}[t]\centering
\includegraphics[width=\columnwidth]{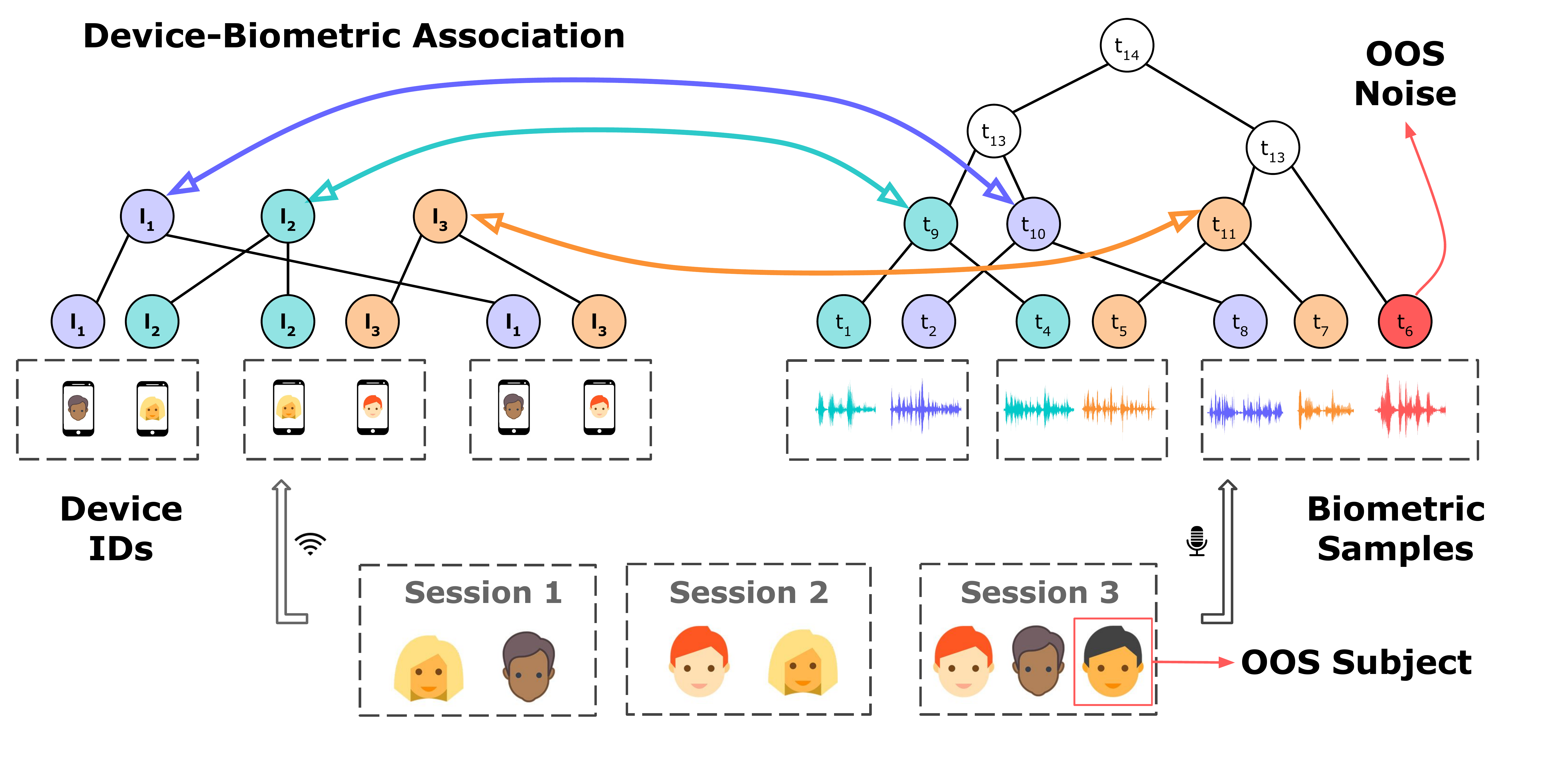}  
\caption{An illustrative example with $3$ victims and $1$ OOS subject. Our presented method inherits the concept of `early stop' and tolerates the noisy biometric samples from OOS subjects before their contamination in merging.}\label{fig:cross_assoc}
\end{narrowFigure}
\section{Evaluation} 
\label{sec:evaluation}

In this section, we conduct experiments on both real-world (refer to Sec.~\ref{sub:data_collection_methodology}) and simulation datasets to quantitatively verify the performance of the presented attack. 

\subsection{Baseline and Performance Measure} 
\label{sub:setup_and_metrics}

\subsubsection{Baseline Methods}
\label{subsub:baseline_method}

Two variants of the proposed association approach are considered: (1) \texttt{Ours (Euc.)} uses an Euclidean distance metric rather than the proposed Dice coefficient to calculate a node score $\mathbf{q}_{assoc_i}$ (see Sec.~\ref{sub:linkage_tree_construction}); (2) \texttt{Naive} is the the intuitive strategy introduced in Sec.~\ref{sub:a_naive_method_and_its_limitation_} that performs clustering and association step sequentially in contrast to our simultaneous approach. 

\subsubsection{Performance Measure - Association Accuracy}
\label{subsub:metirc_accuracy}

From the perspective of \emph{device ID compromise}, to represent how well one can associate physical biometrics with device IDs, we calculate the portion of correctly associated pairs of biometric clusters and device IDs among all victims, denoted as the \textit{association accuracy}. The true identity of a device ID is retrieved from a pre-registered table provided by our volunteered `victims', which maps device MAC addresses to their owners. On the other side, the identity of an individual biometric sample is manually labeled via inspecting. Considering that biometric clusters might be noisy, the true label of a biometric cluster is determined by majority voting. For the ease of readability, we refer to association accuracy as \emph{accuracy} hereafter.

\subsubsection{Performance Measure - Cluster Purity}
\label{subsub:metirc_purity}

From the perspective of a \emph{biometric compromise}, we introduce an additional metric \textit{purity} that concerns sample-level accuracy. Given a pair of correctly associated biometric cluster and MAC address, cluster \textit{purity} describes the ratio of samples that actually belong to the corresponding victim.
This metric stems from the real-world application where pure biometric clusters are more valuable to an attacker for quick identity determination (e.g., majority voting) and launching subsequent attacks, such as biometric spoofing or impersonation attacks. For the ease of readability, we refer cluster purity as \emph{purity} hereafter.

\begin{narrowFigure}[!t]
    \centering
    \begin{subfigure}{\linewidth}
        \centering
        \includegraphics[width=.49\linewidth]{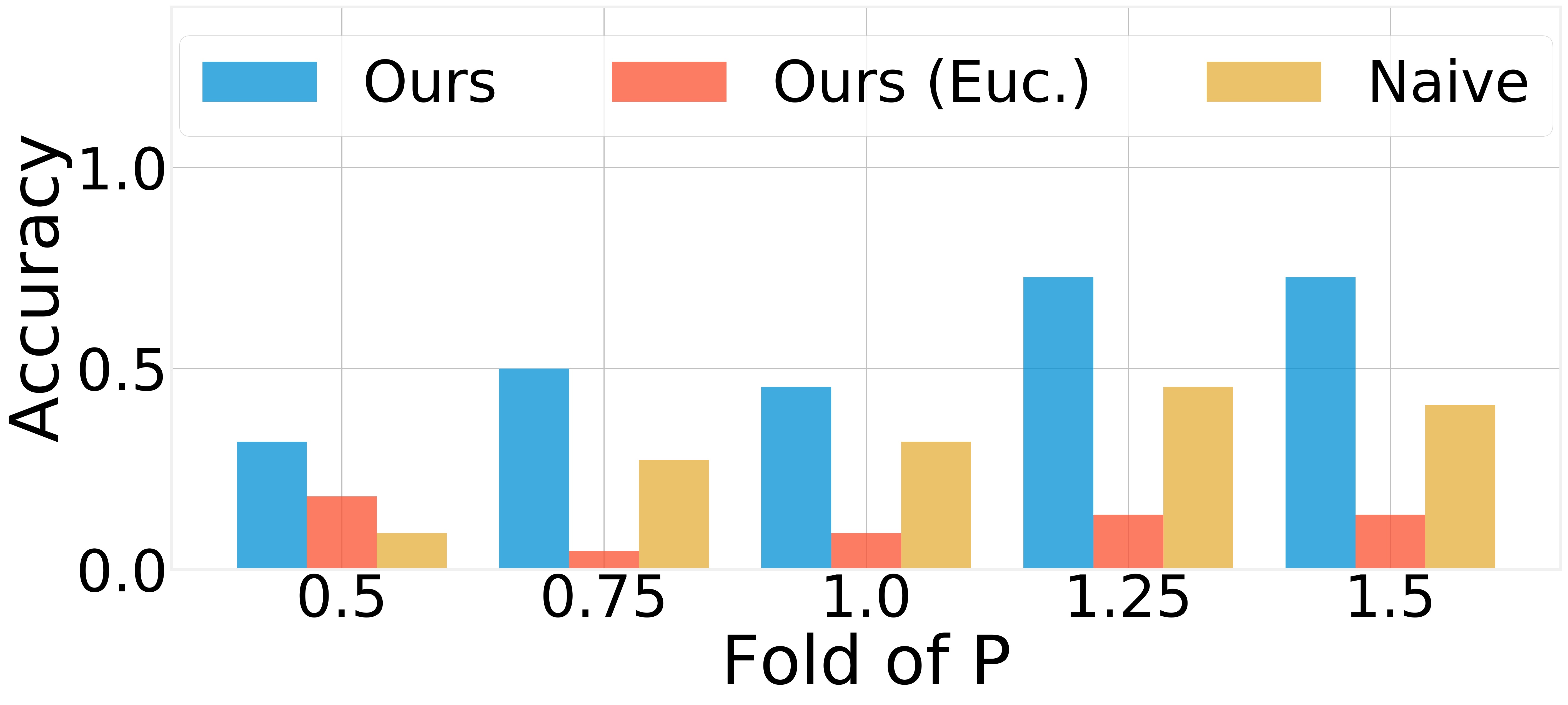}
        \hfill
        \includegraphics[width=.49\linewidth]{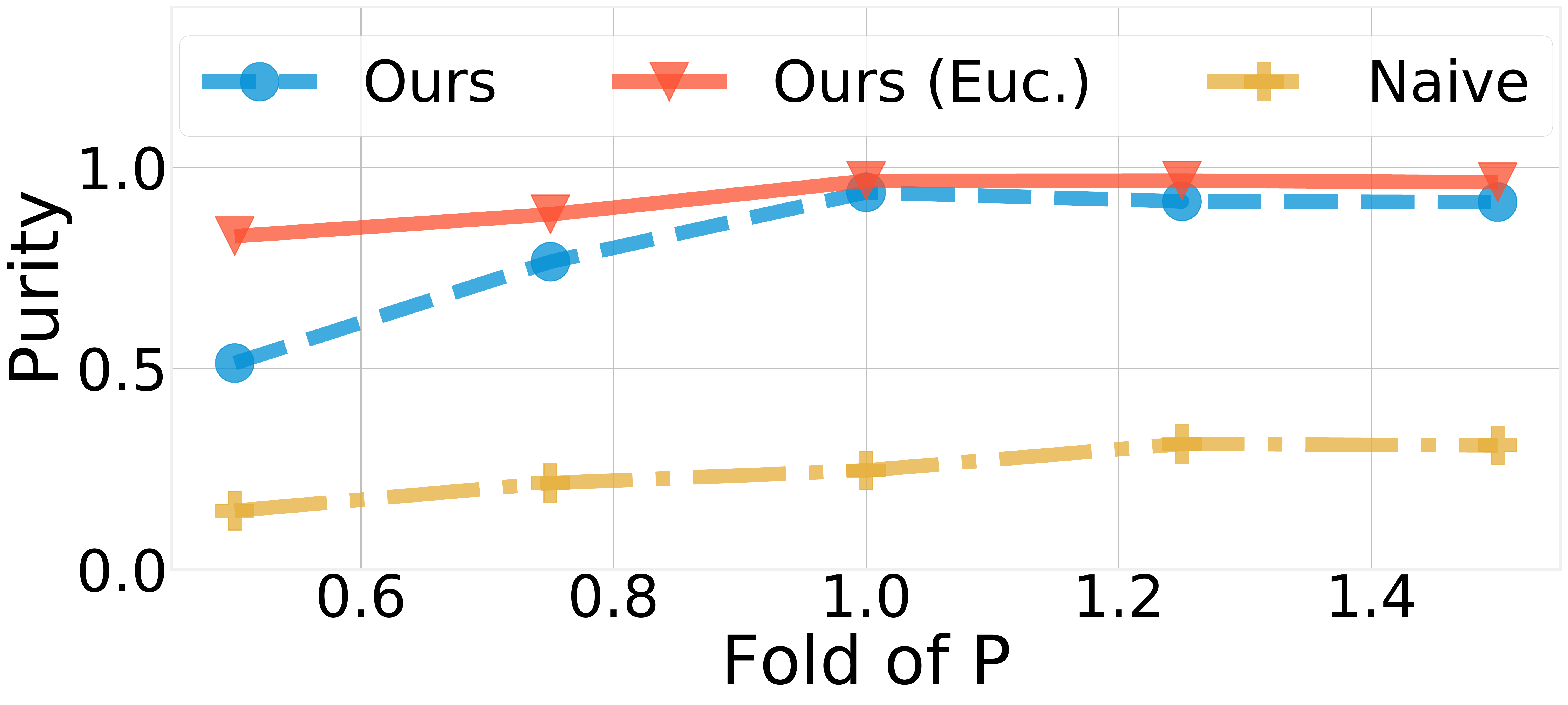}
        \caption{Face-domain.}
    \end{subfigure}
    \vskip\baselineskip
    \begin{subfigure}{\linewidth}
        \centering
        \includegraphics[width=.49\linewidth]{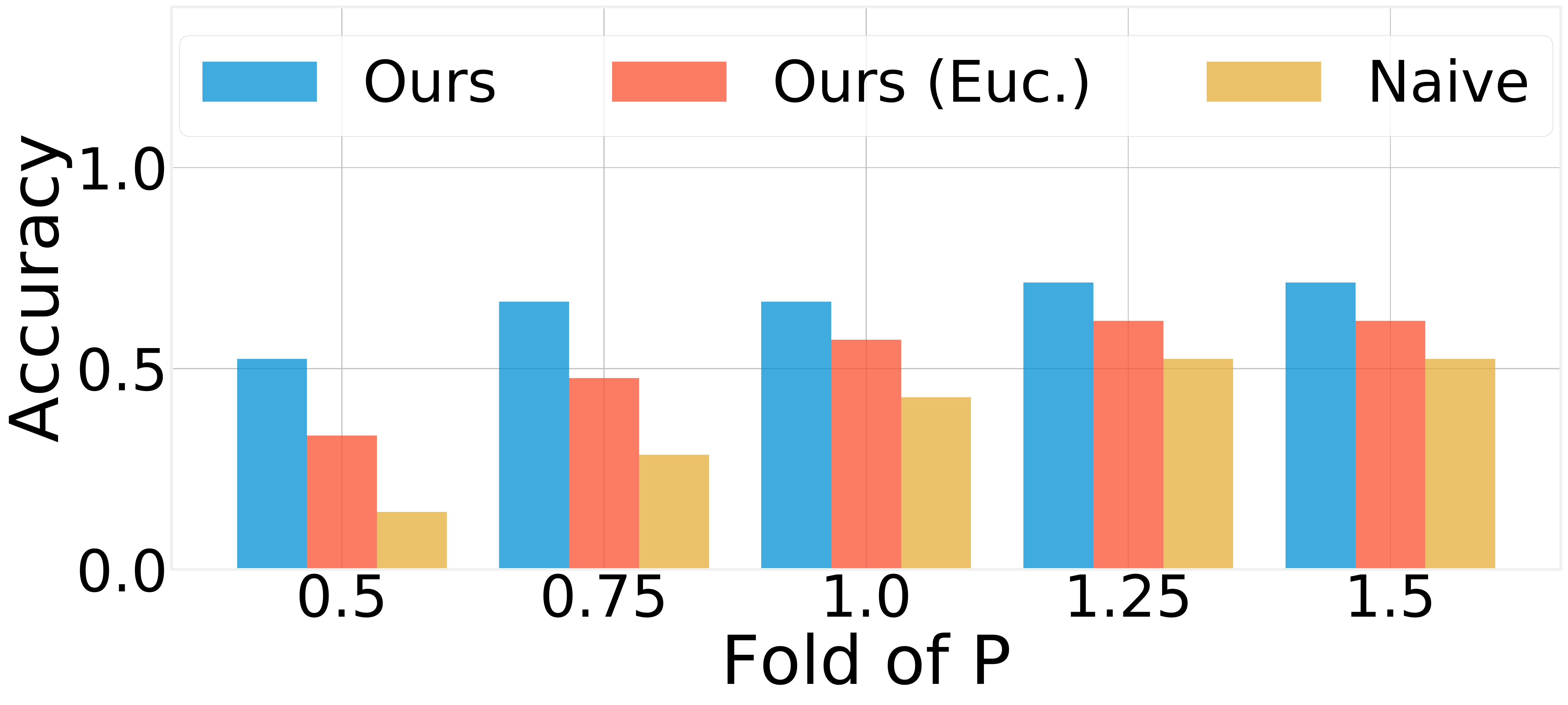}
        \hfill
        \includegraphics[width=.49\linewidth]{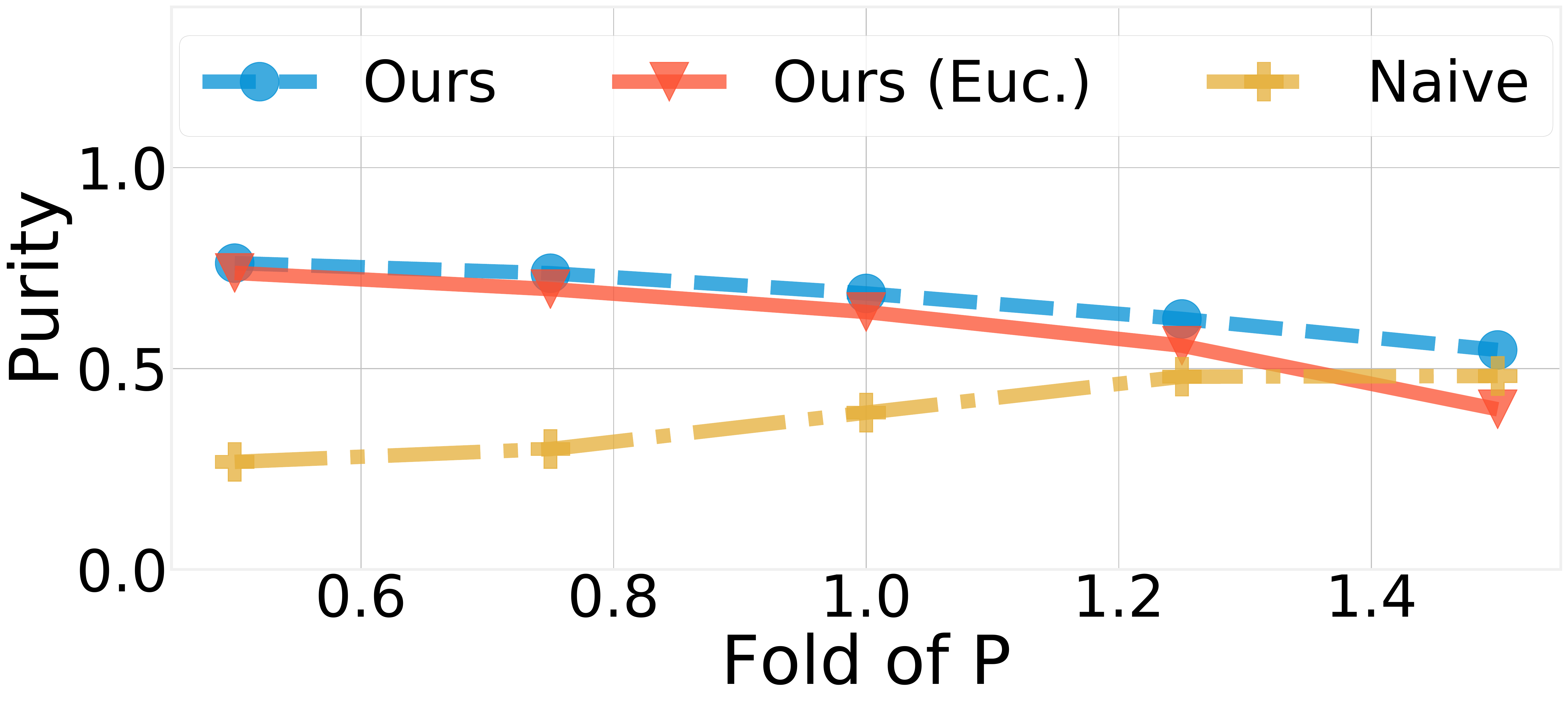}
        \caption{Voice-domain.}
    \end{subfigure}
    \vspace{-8pt}
    \caption{Overall Attack Performance.}
    \vspace{-3pt}
    \label{fig:sensetivity_K}
\end{narrowFigure}

\subsection{Attack Performance} 
\label{sub:attack_performance}

We quantify the cross-modal ID leakage on \textit{RealWorld} datasets. Recall in Sec.\ref{sub:node_selection}, an adversary needs to first make an estimation on the number of potential victims $K$ in the target environment. Such $K$ is generally inaccurate due to OOS disturbances yet deeply interwoven in almost all real-world attacks. Obviously, the attack algorithm needs to be robust against different levels of noises. We therefore report \emph{accuracy} and \emph{purity} under different choices of $K$ to provide a comprehensive view on attack performance.

\noindent \textbf{Setup.}
As a reference, we examine a set of $K$s ranging from 50\% to 150\% of the true number of victims $P$ on the \textit{RealWorld} datasets in both face- and voice-domain. In this experiment, the regularization parameter $\omega$ is set to $0.5$ (refer to Eq.~\ref{eq:score}) and the RSS threshold is set to -55 dBm and -45 dBm for face- and voice-domain respectively. All eavesdropped sessions of our real-world datasets (see Sec.~\ref{sub:data_collection_methodology}) are used in this experiment.

\noindent \textbf{Overall Performance.}
Fig.\ref{fig:sensetivity_K} demonstrates the performance of the presented mechanism and two baselines as $K$ varies. In all cases, a clear winning margin of our method against baselines is observed, with a relatively wide region of $K$ achieving comparable performance. Our approach achieves 72.7\% and 71.4\% \emph{accuracy}, and 93.9\% and 76.2\% \emph{purity} on face and voice domain datasets respectively.

\noindent \textbf{Dataset Comparison.}
It can be seen that the advantage is much more prominent on face-domain. This is due to the fact that faces captured by cameras in the wild are much noisier than utterances captured by recorders as described in Sec.\ref{sec:dataset_and_feasibility}. Consequently, the quality of face clusters is inferior to that of voice clusters.

\noindent \textbf{Comparison with \texttt{Ours (Euc.)}.}
Observing our method (\texttt{Ours}) and its Euclidean variant \texttt{Ours (Euc.)}, it can be observed that Dice coefficient yields a significantly higher \textit{accuracy} than Euclidean distance, especially on face-domain dataset. This is because, compared to Dice, Euclidean distance evenly punishes unmatched and matched instances when searching for similar context vectors of biometric clusters and MAC addresses. However, in our scenario, Dice is more desirable as the matched pairs should be of higher importance over unmatched ones to combat attendance sparsity and observation noise (see Sec.~\ref{sub:linkage_tree_construction}). Consequently, Euclidean distance usually leads to purer clusters if correctly associated, but it is also very error-prone when pairing clusters to MAC addresses. Inspecting \textit{purity}, our method with Dice coefficient is able achieve comparable results with \texttt{Ours (Euc.)} when $K$ is reasonably set.

\noindent \textbf{Comparison with \texttt{Naive}.}
Compared to the \texttt{Naive} approach, the advantage of using simultaneous clustering and association (\texttt{Ours}) is also obvious, especially for \textit{purity}. This can be attributed to the nature of clustering. Commonly, a clustering algorithm requires one to pre-define the number of clusters\footnote{Clustering algorithms without the requirement of number of clusters often require other hyper-parameter specification, e.g., the neighborhood size for DBSCAN \cite{schubert2017dbscan}.}, then separates the dataset accordingly. However, under the scenario of attack, such behavior makes association rely on the guessed number of clusters (i.e., the $K$). However, noises will be scattered inside each cluster and this effect is particularly significant in face-domain. Despite this, our method alleviates the impact of uncertain $K$ and gives best \textit{accuracy} since it only takes effect after identity association. The \textit{purity} of the \texttt{Naive} approach is compromised due to the same reason.

\subsection{Impact Factors} 
\label{sub:impact_of_factors}

We now continue our evaluation on the effects of different impact factors, including (1) WiFi RSS Threshold, (2) Collection Span, (3) regularization parameter $\omega$ and (4) Number of OOS subjects. Notably, for the ease of scaling the number of OOS subjects, the fourth experiment is conducted on two large-scale simulation datasets. Hereafter, unless otherwise stated, all experiments in (1-3) are conducted on 123 sessions in face-domain and 49 sessions in voice-domain. $K$ is set to $1.25$ times of the true number of victims $P$ (see Sec.~\ref{sub:node_selection}) and the regularization parameter $\omega$ is equal to $0.5$ (refer to Sec.~\ref{sub:linkage_tree_construction}). The RSS threshold is set to -$55$ dBm and -$45$ dBm for face- and voice-domain respectively.

\subsubsection{Impact of WiFi RSS Threshold}
\label{subsub:sensitivity_rss}

\begin{narrowFigure}[!t]
    \centering
    \begin{subfigure}{\linewidth}
        \centering
        \includegraphics[width=.49\linewidth]{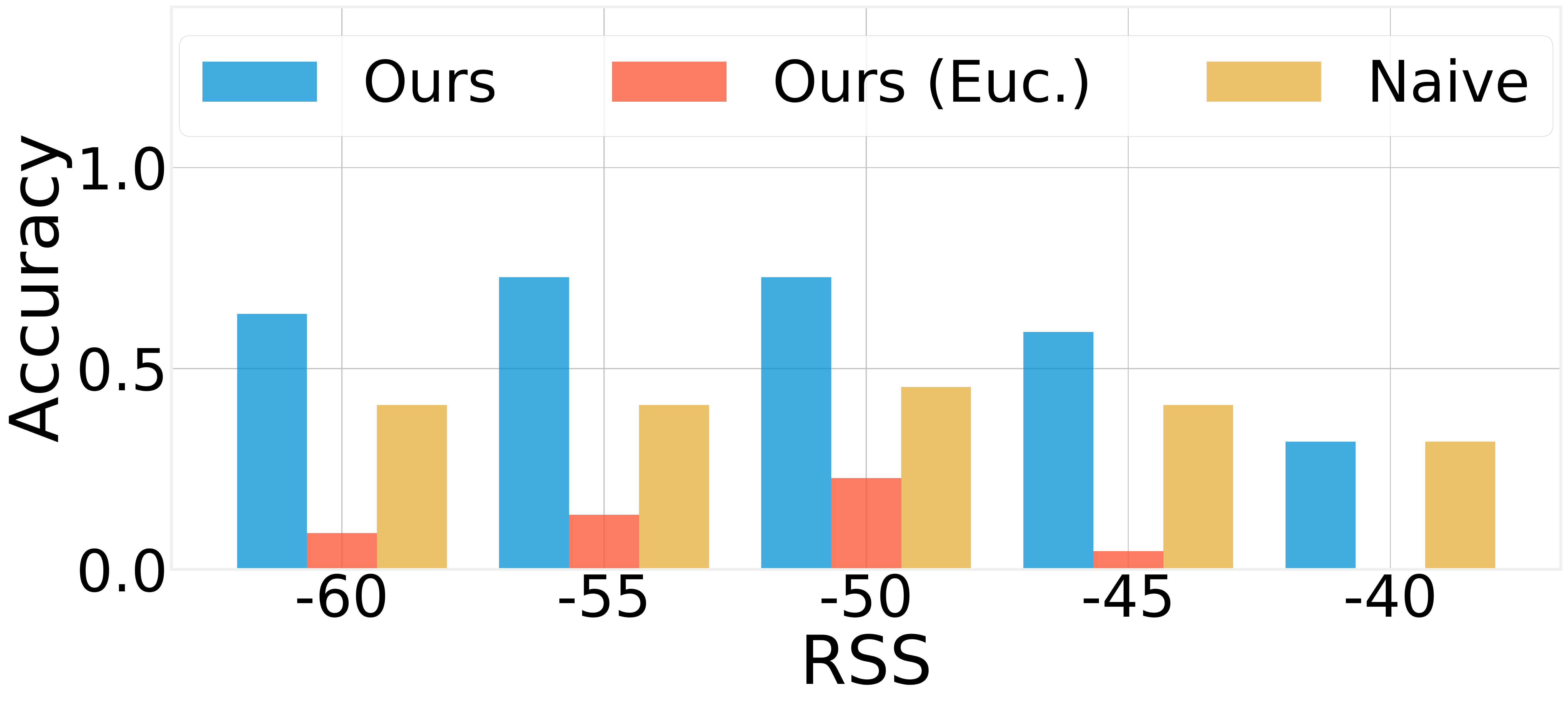}
        \hfill
        \includegraphics[width=.49\linewidth]{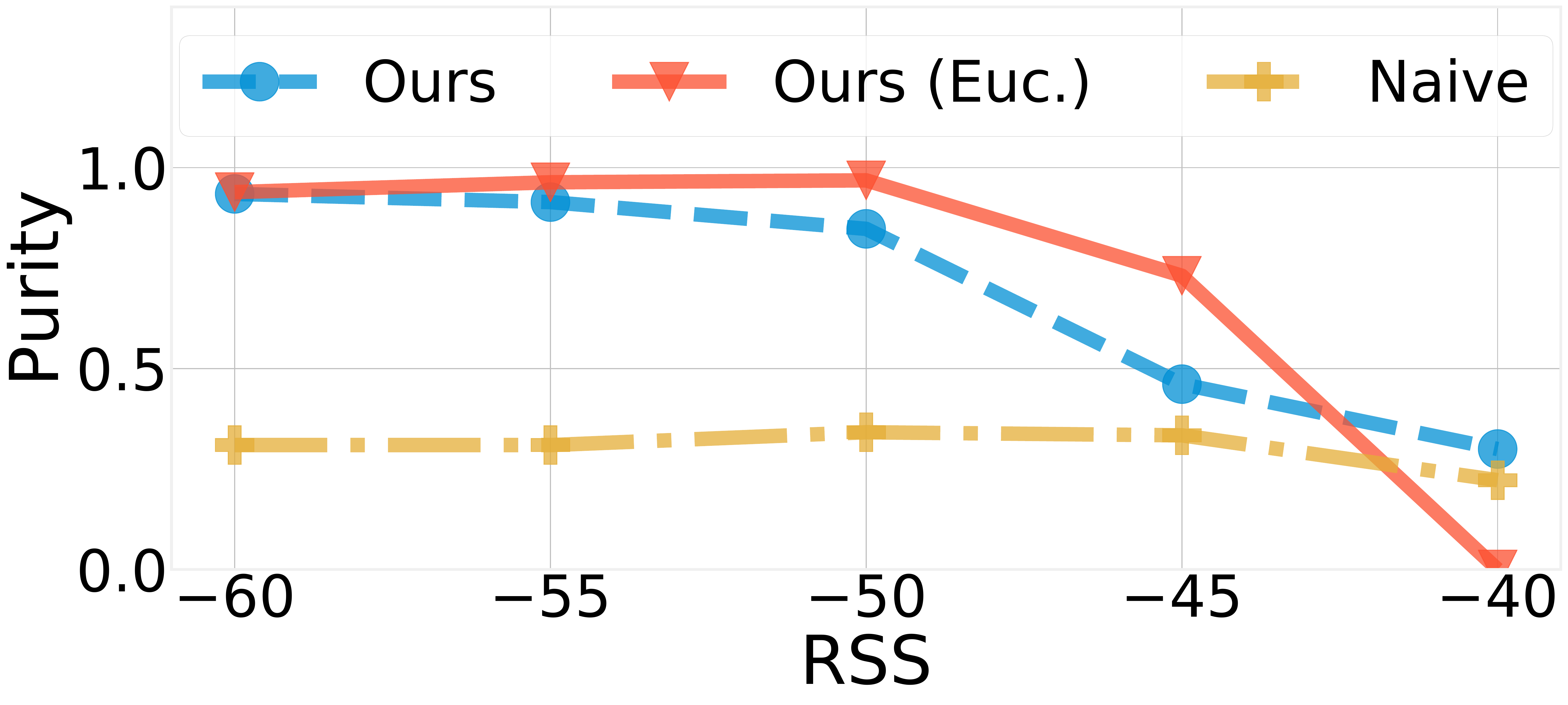}
        \caption{Face-domain.}
    \end{subfigure}
    \vskip\baselineskip
    \begin{subfigure}{\linewidth}
        \centering
        \includegraphics[width=.49\linewidth]{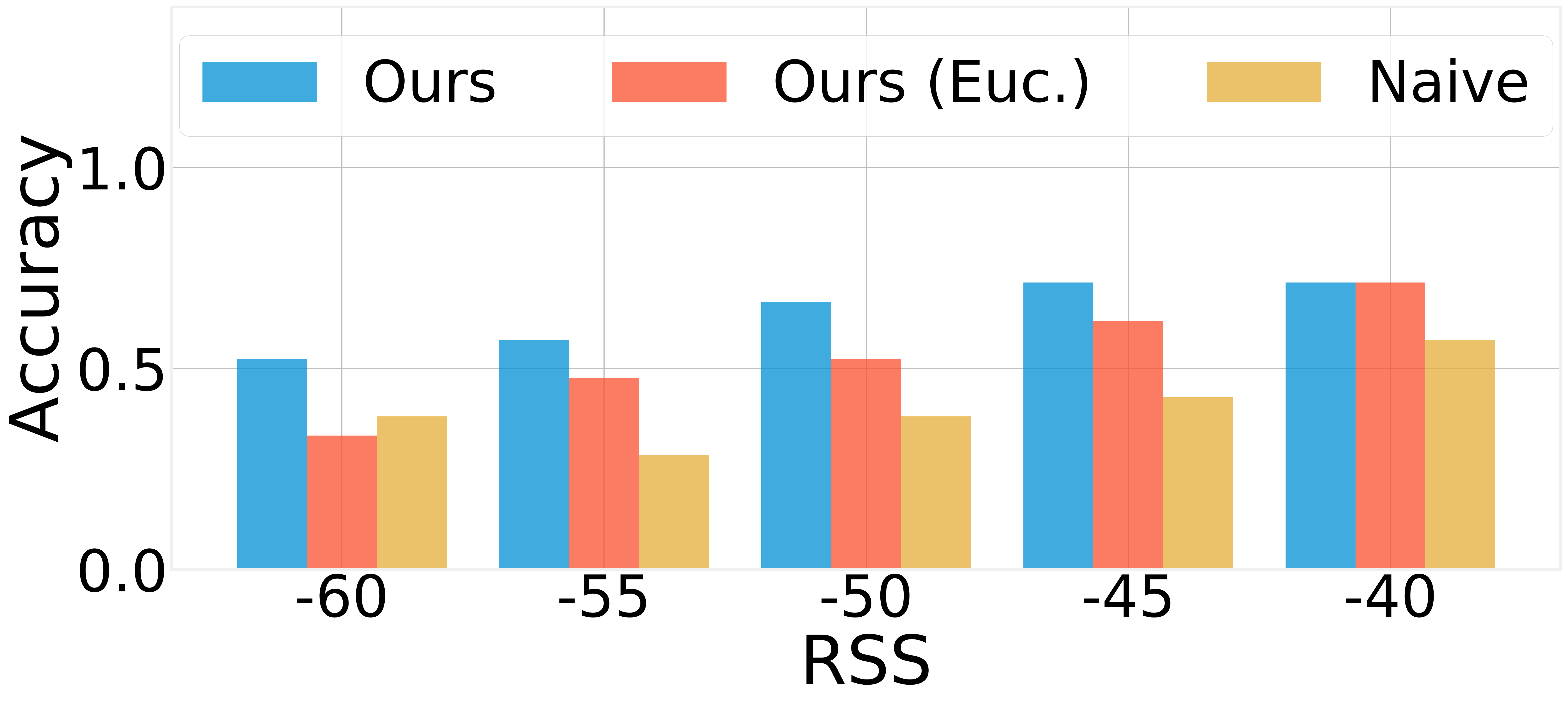}
        \hfill
        \includegraphics[width=.49\linewidth]{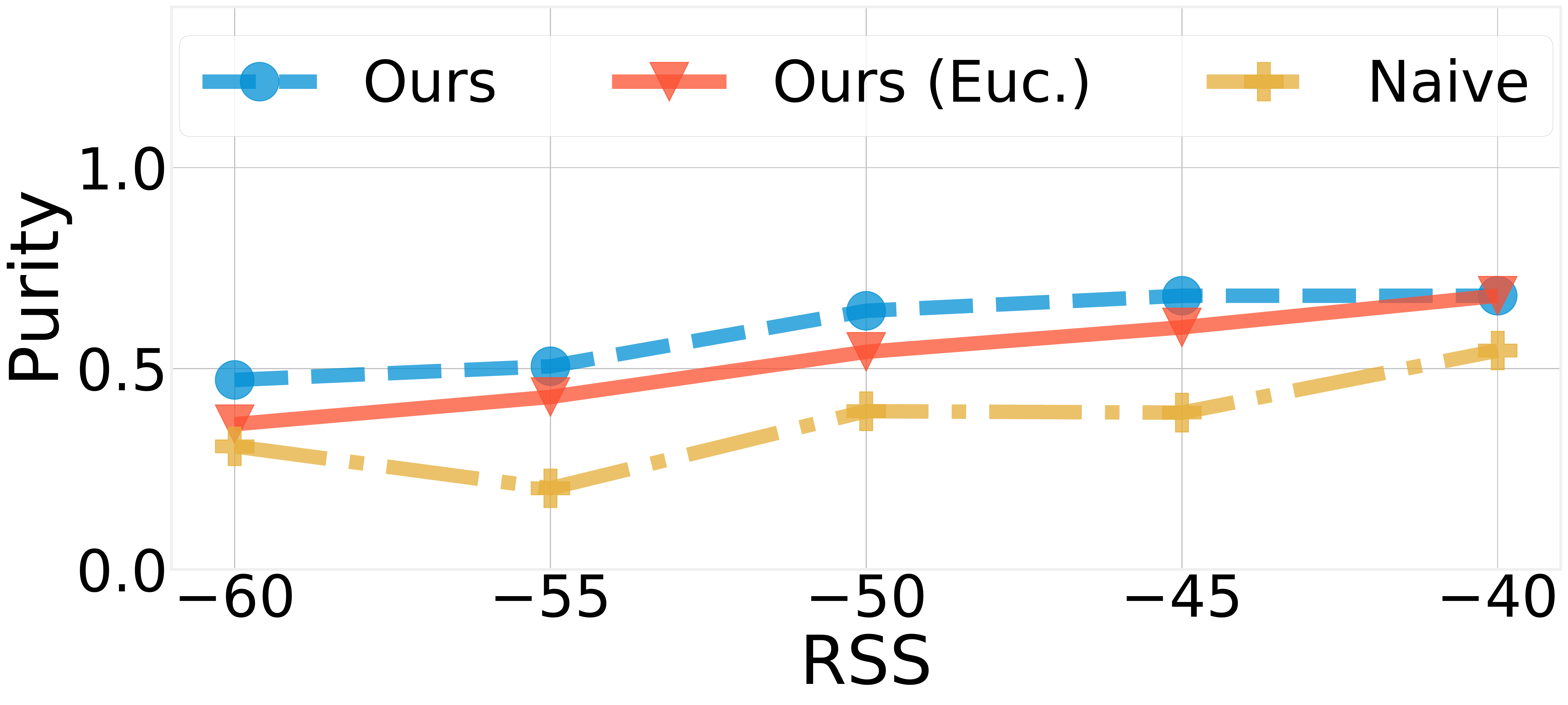}
        \caption{Voice-domain.}
    \end{subfigure}
    \vspace{-8pt}
    \caption{Impact of WiFi RSS threshold on \textit{RealWorld} datasets.}
     \vspace{-3pt}
    \label{fig:sensetivity_rss}
\end{narrowFigure}

As described in Sec.\ref{sub:received_signal_strength_filter}, a critical factor of filtering nuisance MAC addresses is the spatial distance. Leveraging the RSS information contained in sniffed packets, an adversary can determine whether a device is presented in the target environment by comparing with a pre-defined `fence'. Normally, devices closer to the sniffer have stronger RSS. However, without accurate calibration, an adversary can only empirically guess a threshold value based on this fact which results in ambiguity.

As we can see in Fig.\ref{fig:sensetivity_rss}, the influence of varying RSS threshold is global on all three methods. It shows that for face-domain, a rigorous threshold harms association result with a decrease ranging from 22.2\% to 56.2\% in \textit{accuracy}. A drastic drop happens in \textit{purity} when the threshold goes beyond -50 dBm where \texttt{Ours} decreases to 30\% and \texttt{Ours (Euc.)} fails at -40 dBm. In voice-domain, performance degrades when the threshold is relaxed for all three approaches, where \textit{accuracy} shows 26.7\%, 53.3\% and 50.0\% decrements in \texttt{Ours}, \texttt{Ours (Euc.)} and \texttt{Naive} respectively. Similar degradation happens to \textit{purity}. On the other side, the presented approach maintains its advantage in both \textit{accuracy} and \textit{purity}. The discrepancy between face- and voice-domain is mainly due to the difference of locations where faces were captured by the hidden camera deployed in a relative large public office ($60m^2$) while dialogues were mostly recorded in a small meeting room ($25m^2$).

\subsubsection{Impact of Collection Span}
\label{subsub:sensitivity_collection_span}

\begin{figure}[!t]
    \centering
    \begin{subfigure}{\linewidth}
        \centering
        \includegraphics[width=.49\linewidth]{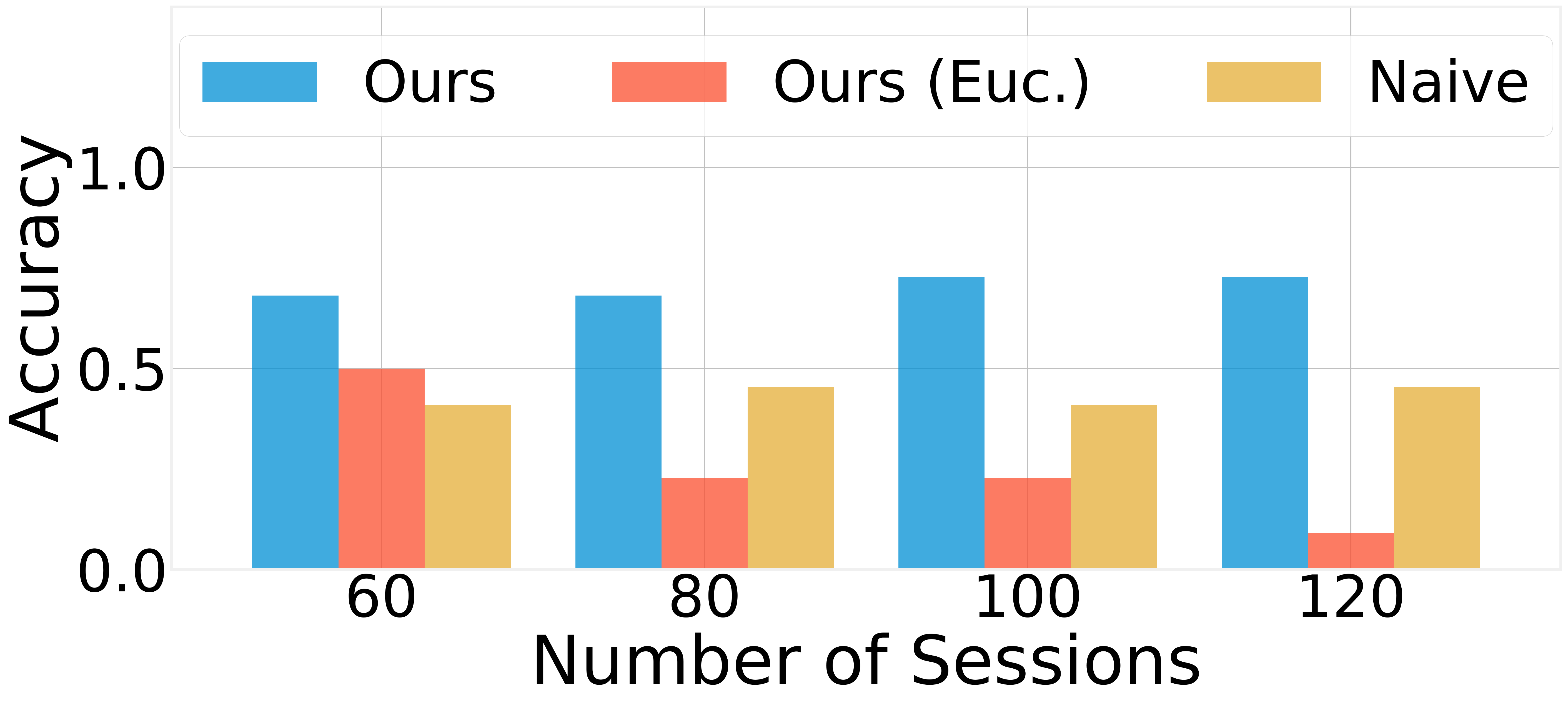}
        \hfill
        \includegraphics[width=.49\linewidth]{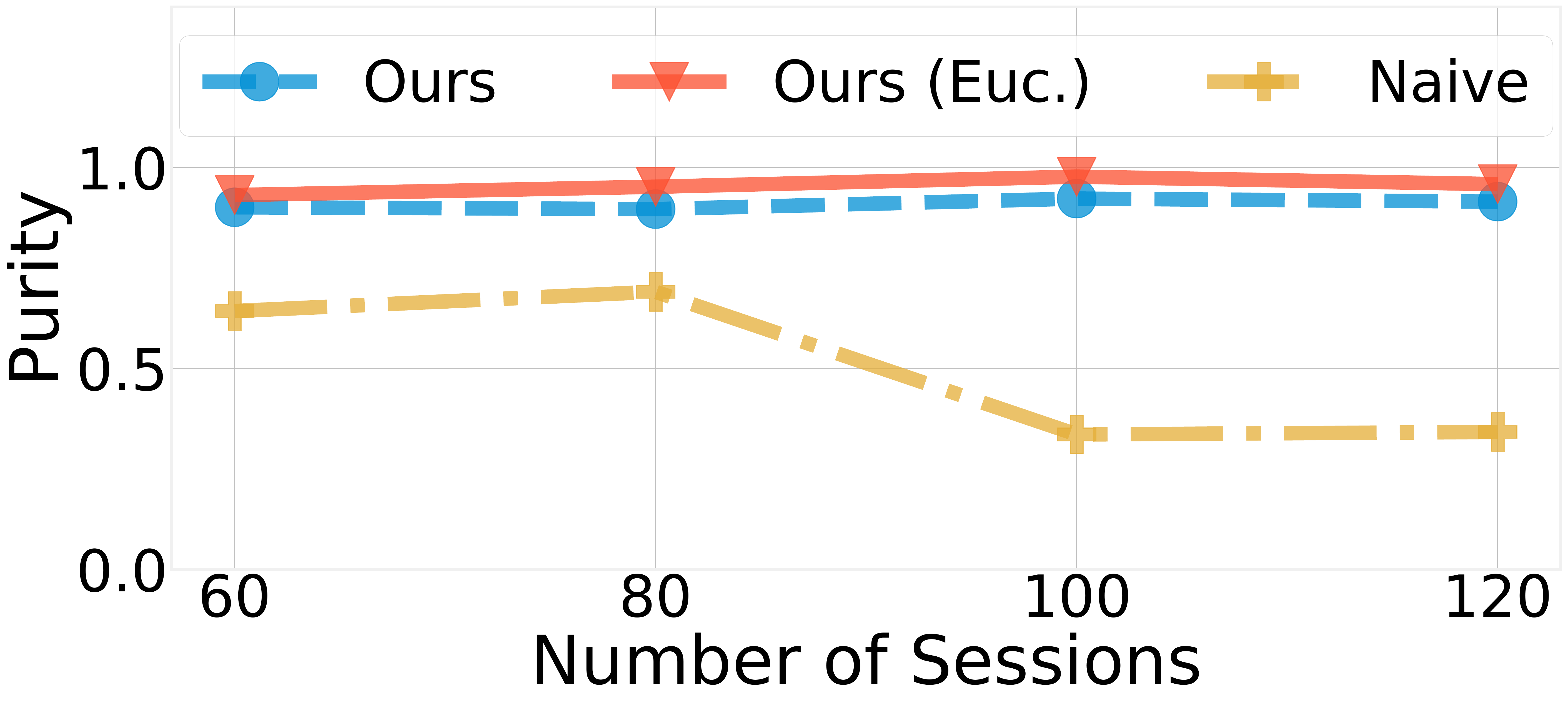}
    \end{subfigure}
    
    \caption{Impact of collection span on face-domain.}
    \vspace{-13pt}
    \label{fig:sensetivity_collection_span}
\end{figure}

\begin{narrowFigure}[!t]
    \centering
    \begin{subfigure}{\linewidth}
        \centering
        \includegraphics[width=.49\linewidth]{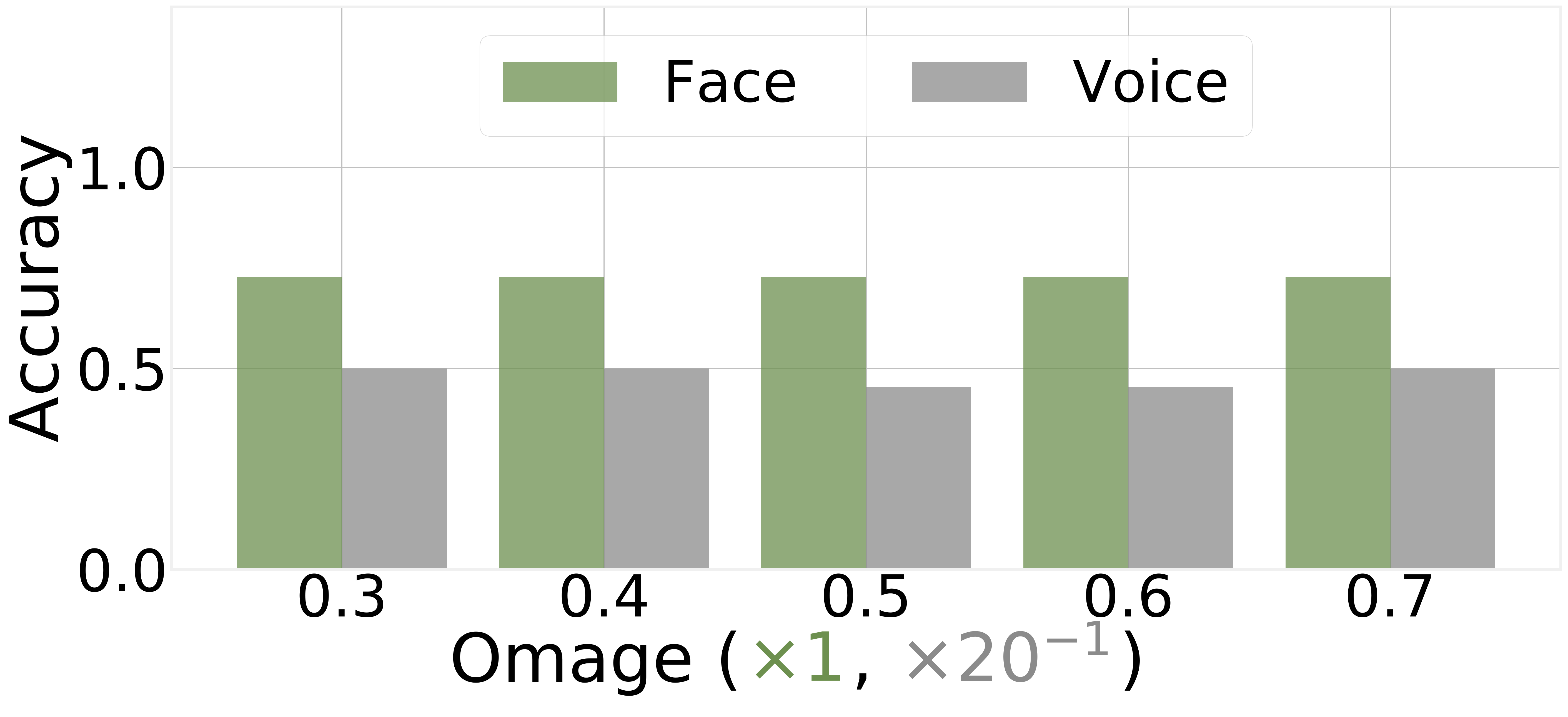}
        \hfill
        \includegraphics[width=.49\linewidth]{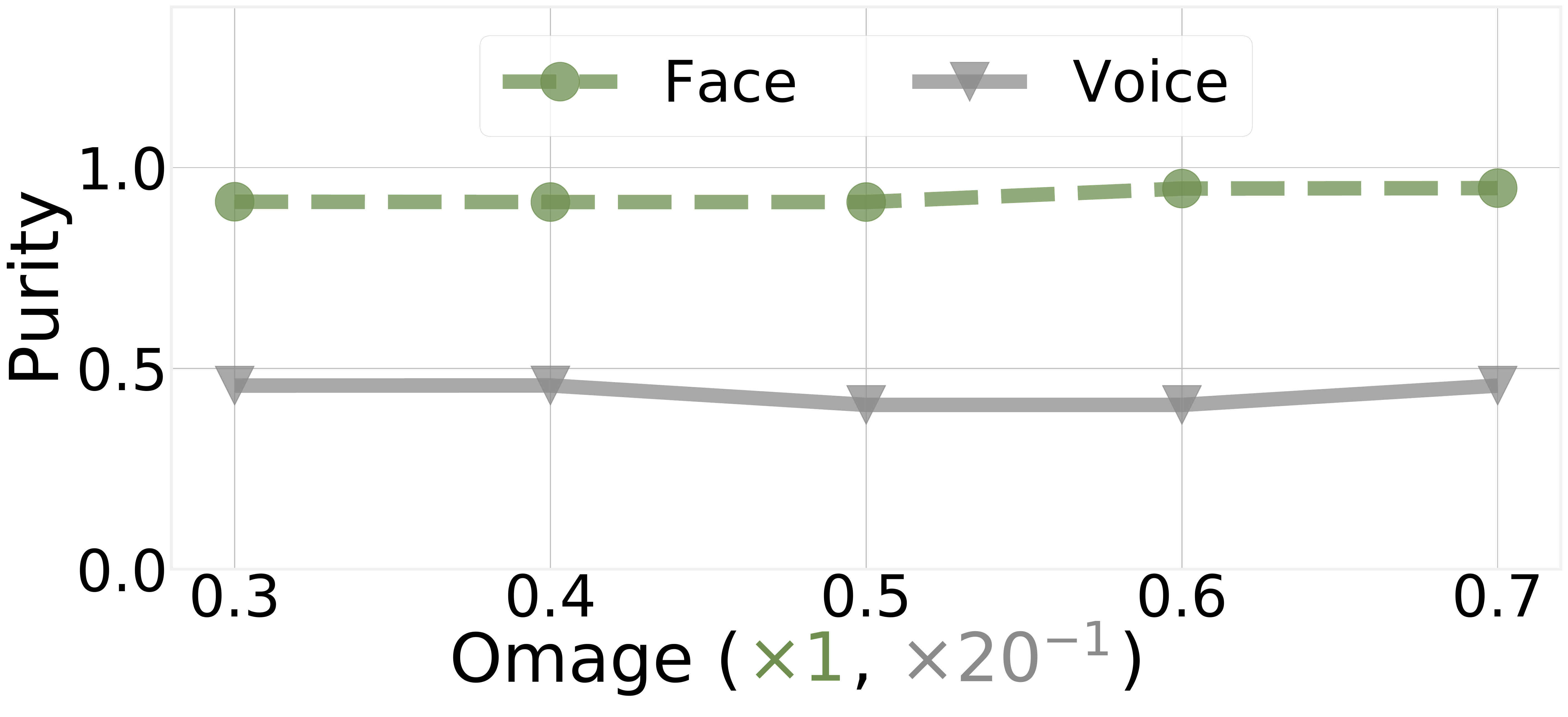}
    \end{subfigure}
    \vspace{-5pt}
    \caption{Impact of omega on \textit{RealWorld} datasets.}
    \label{fig:sensetivity_omega}
\end{narrowFigure}

Another key parameter to consider when executing the attack is the number of eavesdropped sessions, referred to as collection span. Recall that in Sec.\ref{ssub:uniqueness_analysis}, we discussed the prerequisite for victims to be distinguishable. Naturally, more sessions lead to more diverse attendance patterns, indicating that the probability of having inseparable victims is lower. However, it also increases the amount of OSS observations in both biometric samples and device IDs. In this experiment, we study the influence of collection span by gradually decreasing the number of eavesdropped sessions used in association, mimicking the variation from long-term to short-term monitoring. This experiment is not practical on voice-domain dataset since the number of eavesdropped sessions is limited ($49$ sessions) and further decrements in collection span leads nowhere.

Fig.\ref{fig:sensetivity_collection_span} shows that on the face-domain dataset, the presented approach (\texttt{Ours}) can achieve reasonable \textit{accuracy} and \textit{purity} without using all sessions. This observation coincides with our feasibility analysis in Sec.~\ref{ssub:uniqueness_analysis} that attendance uniqueness is widely available even among co-workers.
Note that Dice similarity (\texttt{Ours}) achieves similar purity level with Euclidean distance (\texttt{Ours (Euc.)}) while maintains its advantage in \textit{accuracy}. \texttt{Ours (Euc.)} manages to maintain its high \textit{purity} but the \textit{accuracy} is corrupted due to its non-discriminative constraint on matched and unmatched attendance. Last but not least, there is a drastic drop in \texttt{Naive} approach's \textit{purity} when the number of sessions increases from $80$ to $100$, caused by the increment amount of noise added to clusters. 


\subsubsection{Impact of $\omega$}
\label{subsub:sensitivity_omega}

Recall that in Sec.\ref{sub:linkage_tree_construction}, we introduced a regularization parameter $\omega$ to balance the importance of biometric feature similarity and session attendance similarity. In this experiment, we examine the impact of varying this parameter on the presented approach (\texttt{Ours}). Due to the intrinsic difference between biometric features, $\omega$ takes effect at different scales (i.e., $\times 20^{-1}$ for voice domain against face domain). As shown in Fig.\ref{fig:sensetivity_omega}, our method is robust to this parameter while a slight lift of $3.8\%$ in \textit{purity} can be observed in face-domain when we put more trust on the context vector. Such behavior proves that, even though the features extracted via the pre-trained DNNs are already trustworthy, there is still a gap caused by domain difference. In our approach, context vector can bootstrap association by providing more insights from the eavesdropped sessions. As a comparison, since recorded vocal segments are much less noisy in voice-domain, biometric clusters become more trustworthy.   

\subsubsection{Impact of Number of OOS Subjects}
\label{subsub:sensitivity_oos}

\begin{narrowFigure}[!t]
    \centering
    \begin{subfigure}{\linewidth}
        \centering
        \includegraphics[width=.49\linewidth]{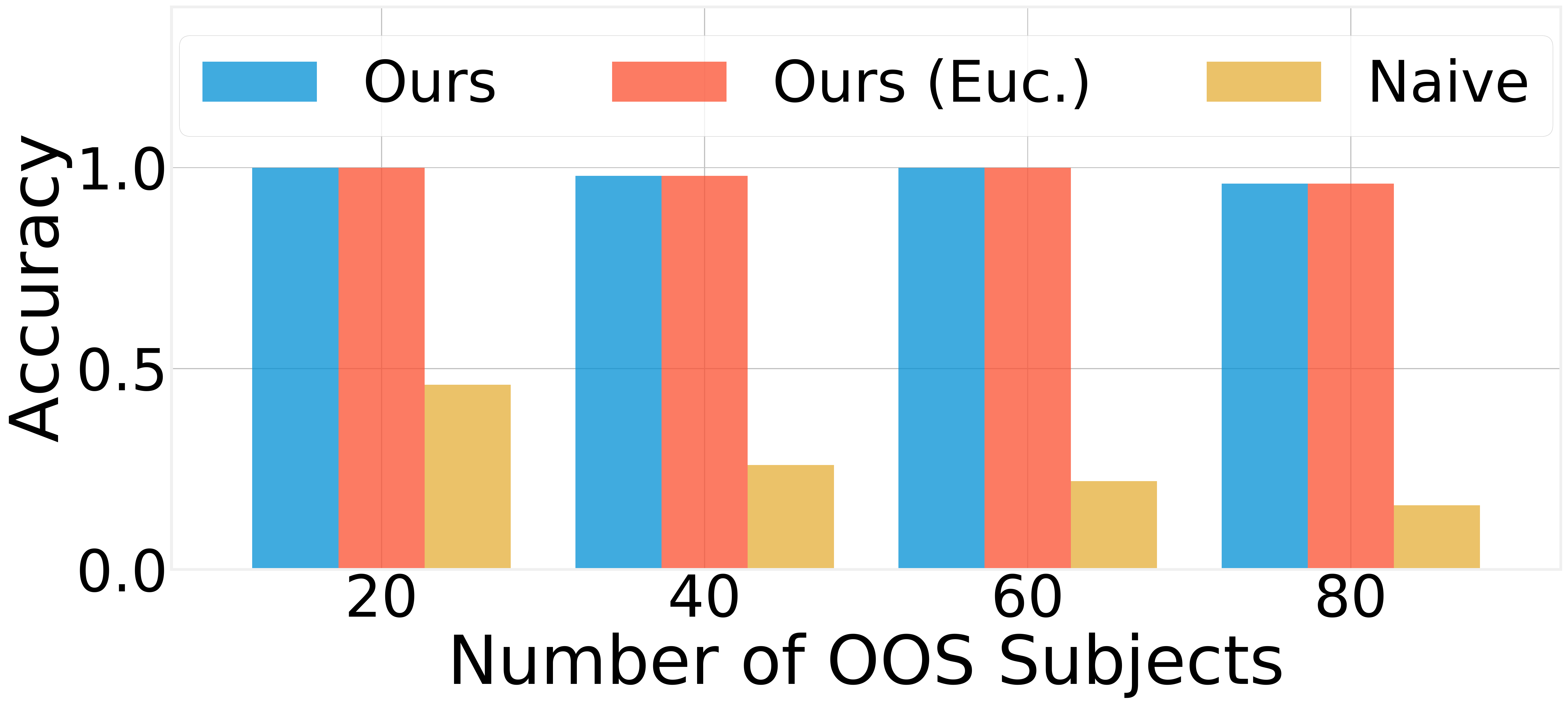}
        \hfill
        \includegraphics[width=.49\linewidth]{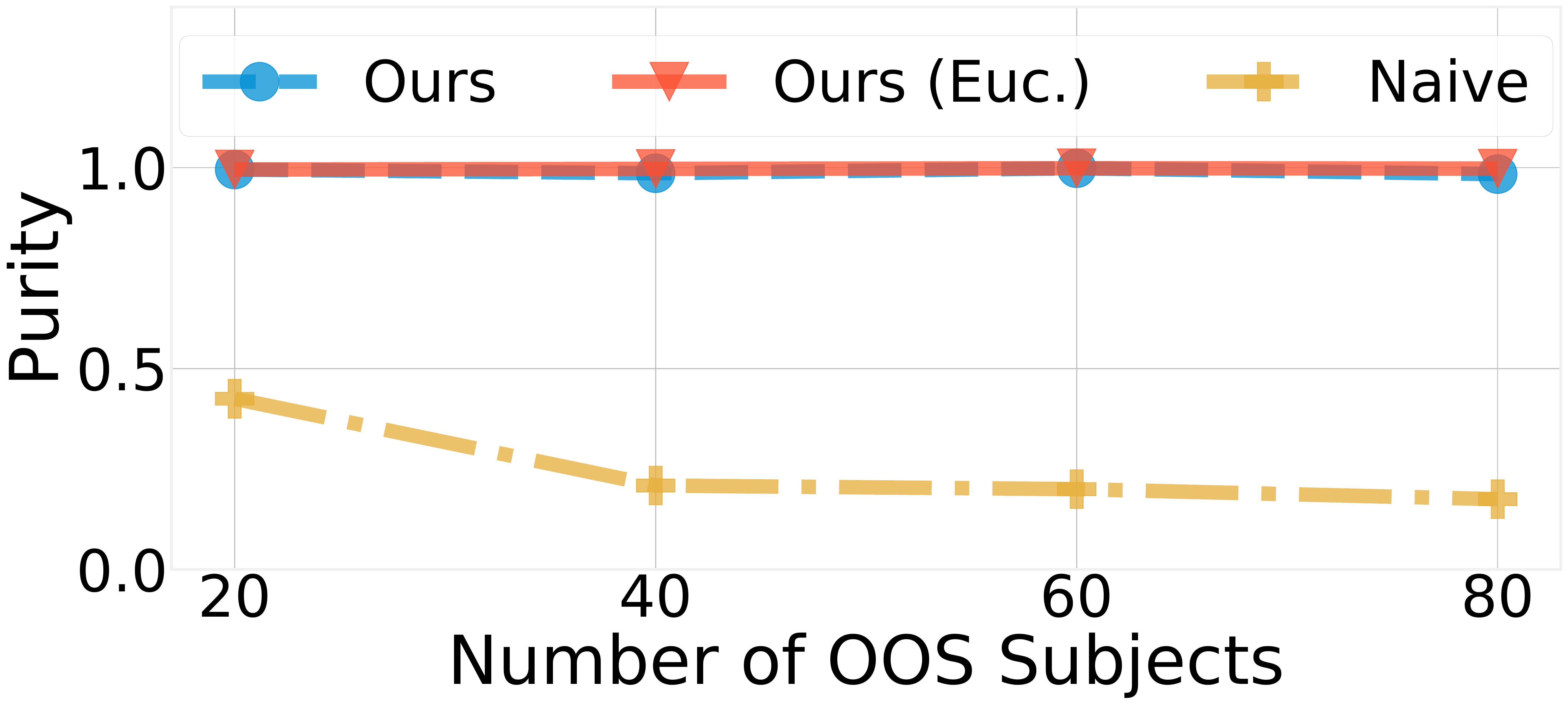}
        \caption{Face-domain.}
    \end{subfigure}
    \vskip\baselineskip
    \begin{subfigure}{\linewidth}
        \centering
        \includegraphics[width=.49\linewidth]{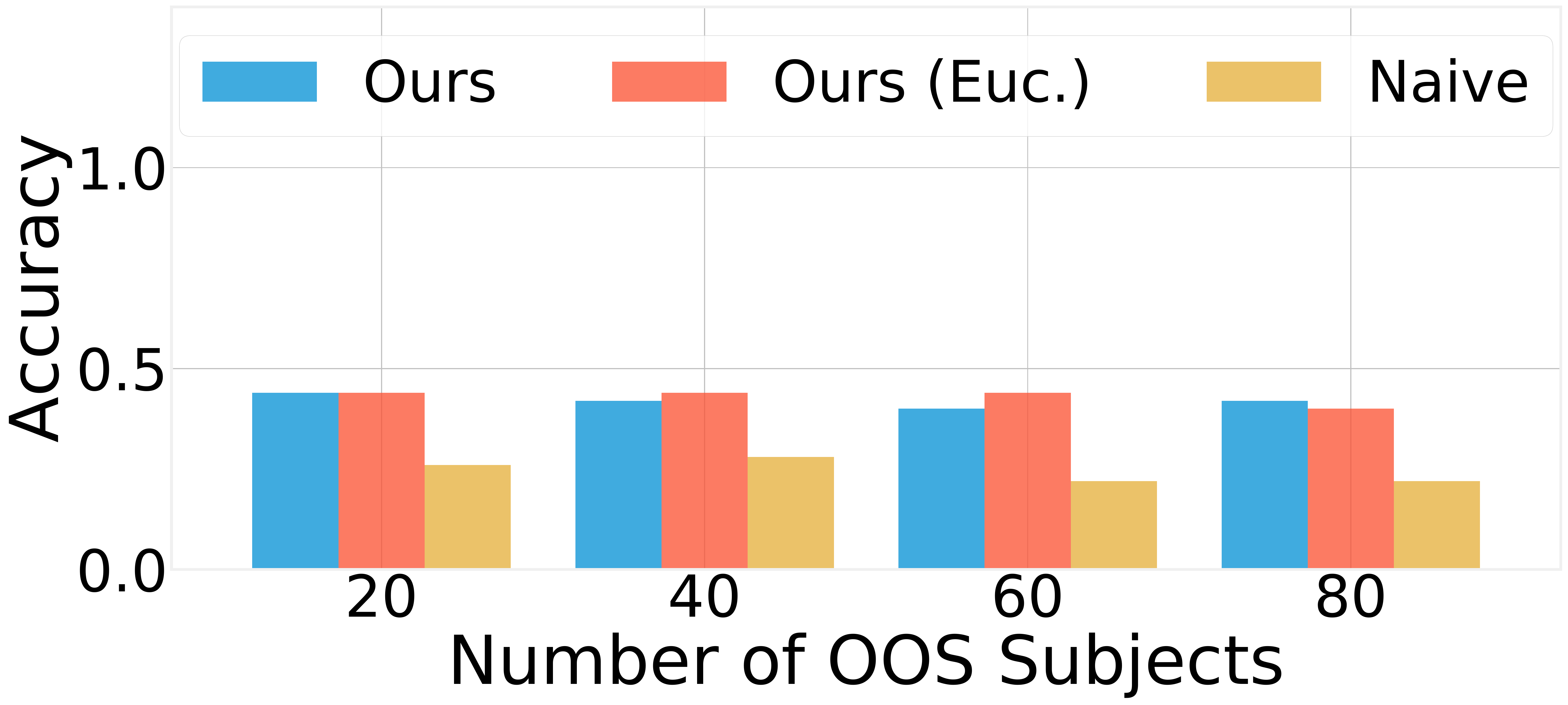}
        \hfill
        \includegraphics[width=.49\linewidth]{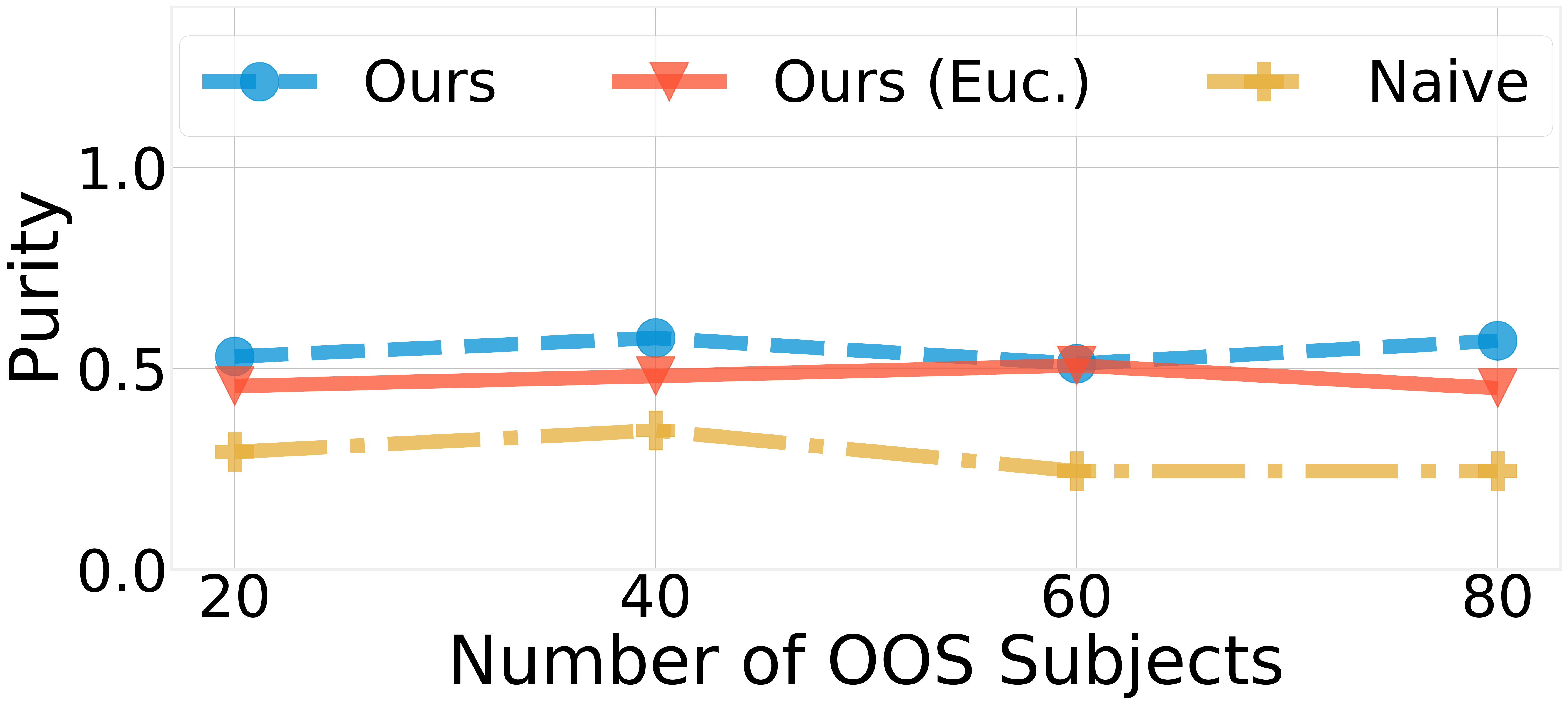}
        \caption{Voice-domain.}
    \end{subfigure}
     \vspace{-8pt}
    \caption{Impact of OOS subjects on \textit{Simulation} datasets.}
     \vspace{-3pt}
    \label{fig:sensetivity_OOS}
\end{narrowFigure}

In this experiment, we examine the robustness of the demonstrated approach where the number of OOS subjects increases, corresponding to different levels of disturbance in biometric observations. To this end, we composed a \textit{Simulation} dataset for each scenario to flexibly adjust this parameter. The \textit{Simulation} dataset was synthesized from publicly available datasets, namely VGGFace2 \cite{Cao18} for face-domain and VoxCeleb2 \cite{Chung18b} for voice-domain. 
Both simulation datasets contain $100$ sessions with $50$ victims and the number of OOS subjects varies from $20$ to $80$ with a step size of $20$. By randomly allocating biometric samples and pseudo device IDs, the datasets contain $26,901$ facial images and $10,100$ voice segments.

As we can see in Fig.\ref{fig:sensetivity_OOS}, \texttt{Ours} still outperforms baselines and yields strong robustness against different levels of OOS disturbance.  
Respectively, in face- and voice-domain, the standard deviations of \textit{accuracy} are $0.02$ and $0.01$, while the standard deviations of \textit{purity} are less than $0.005$ and $0.02$. This result implies that an attacker can leverage our approach to robustly compromise identities in public area, such as public shops with many OOS subjects.


\section{Related Work} 
\label{sec:related_work}

\noindent \textbf{Linkage Attack.} 
The linkage attack is a class of attacks that breaks k-anonymity and reidentifies users \cite{sweeney2002k}. In such an attack, the adversary collects auxiliary information about a specific individual from multiple data sources and then combine that data to form a whole picture about their target, which is often an individual's personally identifiable information. 
Linkage attacks have a long history and can be dated back to the famous de-anonymization of a Massachusetts hospital discharge database by joining it with a public voter database \cite{sweeney1997weaving}. Recent studies demonstrated that linkage attacks are able to infer privacy across different scenarios, including movie ratings forums \cite{narayanan2008robust,frankowski2006you}, crowdsourced sensor/mobility data \cite{lane2012feasibility,jia2017pad,xu2017trajectory} and social networks \cite{zheleva2009join,mislove2010you}. Clearly, our work draws a similar flavor with the linkage attack on utilizing entity uniqueness. But in contrast, our work takes one step forward and formulate the attack problem in the context of IoT-rich environments. To our best knowledge, our work is the first one that systemically studies the vulnerable linkage between biometrics and device IDs.

\noindent \textbf{Side Channel Attacks.}
The side channel attack is proven to be widely powerful in practice. By measuring side channel information, the attacker is able to recover other information that is often sensitive \cite{joy2011side}. There mainly two types of side channel attacks. First, it is the side channel attack based on a single field. Such observed side channel attacks in literature include timing attacks \cite{dhem1998practical}, power analysis attacks \cite{amiel2007passive}, electromagnetic analysis attacks \cite{real2009enhancing}, acoustic attacks \cite{backes2010acoustic} and traffic analysis attacks \cite{kadloor2010low}. The second type is known as a combined side channels attack, which could improve the accuracy of existing attacks or cope with more sophisticated attack scenarios. Some research recently started to explore the attacks from combined side channels \cite{spreitzer2017systematic,lipp2016armageddon,lin2014screenmilker,fan2018constructive}. However, these work only focus on a single domain, which has limitations in employing the information from both sides. In contrast, our system is the first to explore the privacy issue of identity leakage between physical (e.g., user biometrics) and digital domains (e.g., smart devices ID).


\noindent \textbf{Cross-modality Association.}
Technically, our presented cross-modal association method is related to data association methods. Given a track of sensor readings, data association aims to figure out inter-frame correspondences between them. Data association is widely used in radar systems, when tracking blips on a radar screen \cite{blackman1986multiple}, as well as in object monitoring of surveillance systems \cite{hu2004survey}. When it comes to the cross-modal association, research attention is limited and all dedicated to location tracking of humans \cite{teng2014ev,alahi2015rgb,papaioannou2015accurate}. These methods heavily rely on the hypothesis that both sensor modalities are observing evolving state spaces matched precisely in the temporal domain. However, this hypothesis is invalid in our case, since eavesdropping a MAC address does not imply that someone will be speaking at that exact instant. On the other side, association methods are also proposed for autonomous human identification \cite{lu2019autonomous,lu2017scan,lu2019autonomous-b}. However, owing to the purpose of learning, these methods require a \emph{known mapping} between users and their MAC addresses which are \emph{unknowns} to de-anonymize in our context.


\section{Discussion} 
\label{sec:discussion}
 

\subsection{Potential Mitigation Techniques} 
\label{sub:potential_mitigation_techniques}

Since the presented attack is cross-modal, we discuss possible mitigation from both sides of device IDs and biometrics.

\subsubsection{Mitigation from the side of Device ID} 
\label{ssub:mitigation_from_the_side_of_device_id}

In this work, the eavesdropped device IDs (MAC addresses) are obtained from the WiFi sniffing module. Perhaps the most intuitive mitigation against is MAC addresses randomization, a perturbation approach that regularly changes a device's MAC address in \emph{probe request frames} and hence adds more noise to the sniffed data.
However, as pointed out in \cite{martin2017study,vanhoef2016mac}, random MAC addresses, on its own, does not guarantee ID privacy. As randomization is not imposed on the sniffed \emph{association frames} and \emph{transmission frames}, an attacker is still able to capture the true/global MAC addresses of devices. Our analysis on random MAC addresses filter in Sec.~\ref{sec:device_filtering} also validates this observation. Applying randomization to association and transmission frames is difficult as their response mechanisms are managed within the 802.11 chipset, instead of the operating system, implying the only way is to develop a firmware patch that has to be distributed by manufactures. Moreover, even without WiFi sniffing, an attacker may also attain device ID by sniffing on connectivity without using any randomization, e.g., ZigBee \cite{zigbee_sniffing}. Injecting noise to the context vectors by turning off connectivity every once in a while seems a plausible mitigation as well. However, considering the session duration in Tab.~\ref{tab:dataset_description}, an effective noise injection would either cause network interruptions if configured by device manufacturer or require constant user effort when configured manually. 

\subsubsection{Mitigation from the side of Biometrics} 
\label{ssub:mitigation_from_the_side_of_biometrics}

Mitigation solutions on the side of biometrics are boiled down to the long-standing issues of spy cameras and microphones. With the maturity of manufacturing technology and advent in energy harvesting, these hidden devices nowadays can be made smaller and batteryless, making long-term eavesdropping widely available. A promising way is to use e-devices recognition methods such as \cite{li:sensys2018,connolly2008x,hertenstein2011detection}. However, these detection solutions usually require scanning at the entrance before making sure of the space clear, which is heavily dependent on user's cooperation and awareness of hidden devices. 

\subsection{Attack Limitation} 
\label{sub:attack_limitation}

The association between device IDs and physical biometrics is largely based on the discriminative patterns of individuals' session attendance. It is expected that our presented cross-modal association would become ineffective when these patterns become ambiguous and target victims will `Hide in a Crowd'. Such pattern ambiguity can be attributed to over-few eavesdropping sessions, or tricky attendance scenarios where two or more users always appear together during eavesdropping.

\section{Conclusion} 
\label{sec:conclusion}

In this work, we describe a new type of privacy leakage under multi-modal attack. We systemically validated that co-located physical biometrics and device IDs are side channels to each other, which can significantly facilitate the malicious inference of identity thefts. A cross-modal association approach is presented by which the adversary can automatically pair victims with their digital devices and harvest their biometric clusters at the same time, even under complicated sensing disturbances in the real world. With this strategy, the adversary can comprehensively profile victims in a target environment and launch more personal attacks. Experimental results show that our approach is feasible in two real-world scenarios, where face images and voice segments are captured and associated with device MAC addresses. We discussed the limitation of the demonstrated attack vector as well as potential countermeasures against it. 
By raising awareness, we invite the community and IoT manufacturers to look into this new privacy threat.

\bibliographystyle{ACM-Reference-Format}
\bibliography{reference}


\end{document}